\documentclass[epj]{svjour}
\usepackage{graphics}
\usepackage{amsmath,amssymb,cite,color}
\newcommand{\be}{\begin{equation}}
\newcommand{\ee}{\end{equation}}
\newcommand{\ba}{\begin{array}}
\newcommand{\ea}{\end{array}}
\newcommand{\bea}{\begin{eqnarray}}
\newcommand{\eea}{\end{eqnarray}}

\newcommand{\no}{\nonumber}

\newcommand{\cL}{\mathcal{L}}

\newcommand{\cB}{\mathcal{B}}

\newcommand{\BRBsmumu}{\cB(B_s \to \mu^+ \mu^-)}

\newcommand{\rhob}{\bar\varrho}
\newcommand{\etab}{\bar\eta}
\newcommand{\fb}{fb$^{-1}$}
\begin{document}
\title{Status of indirect searches for New Physics with heavy flavour decays after the initial LHC run}
\author{Gino Isidori\inst{1,2} \and Frederic Teubert\inst{2}
}                     
\authorrunning{G.~Isidori \& F.~Teubert}
\titlerunning{Status of indirect searches for New Physics with heavy flavour decays}
\institute{INFN, Laboratori Nazionali di Frascati, Via E.~Fermi 40, I-00044 Frascati, Italy \and CERN, PH Department, Geneva 23, CH-1211.}
\date{Received: date / Revised version: date}
%
\abstract{
We present a status report on the indirect searches for New Physics performed by means of heavy flavour decays. Particular attention is devoted to the 
recent experimental results in $B$ and charm physics obtained by the LHC experiments. The implications of these results for  physics beyond the Standard Model are discussed both in general terms and by means of a few specific examples. 
\PACS{
      {PACS-key}{describing text of that key}   \and
      {PACS-key}{describing text of that key}
     } 
} 
\maketitle
\section{Introduction}
\label{sec:1}

According to the Standard Model (SM) of  fundamental interactions the basic constituents of matter can be grouped into three {\em families}, or {\em flavours}, of quarks and leptons. The four fermions within each family have different combinations of strong, weak,  and electromagnetic charges, that determine completely their fundamental interactions but for gravity. Ordinary matter consists essentially of particles of the first family. As far as we know, quarks and leptons of the second and third family are identical copies of those in the first family but for different, heavier, masses. The heavy quarks and charged leptons are unstable states that can be produced in high-energy collisions and that decay very fast into the lighter fermions of the first family. Why we have three almost identical replica of quarks and leptons, and which is the origin of their different masses, is one of the big open questions in fundamental physics.

In the limit of unbroken electroweak symmetry none of the basic constituents of matter could have a non-vanishing mass. The problem of quark and lepton masses is therefore intimately related to one of the other open key questions in particle physics: which is the mechanism behind the breaking of the electroweak symmetry, or which is the mechanism responsible for the non-vanishing masses of the weak gauge bosons.
Within the SM these two problems are both addressed by the Higgs mechanism: the masses of quarks and leptons, as well as the masses of $W$ and $Z$ bosons, are the result of the interaction of these basic fields  with a new type of field, the Higgs scalar field, whose ground state breaks spontaneously the electroweak symmetry.

The recent observation by the ATLAS~\cite{Aad:2012tfa} and CMS~\cite{Chatrchyan:2012ufa}
 experiments of a new state compatible with the properties of the Higgs boson 
(or the spin-0 excitation of the Higgs field) has significantly reinforced the evidences in favour of the Higgs mechanism and the validity of the SM. However, we have also clear indications that this theory is not complete: the phenomenon of 
neutrino oscillations and the evidence for dark matter 
cannot be explained within the SM.  The SM is also affected by a serious theoretical problem 
because of the instability of the Higgs sector under quantum corrections.
We have not yet enough information to unambiguously determine how this theory should be extended.
However, several well-motivated proposals point toward the existence of new degrees of freedom in the TeV 
range, possibly accessible at the high-$p_T$ experiments at the LHC. 

The description of quark and lepton masses in terms of the Higgs mechanism is particularly unsatisfactory since the corresponding interactions 
between fermion and Higgs fields are not controlled by any symmetry principle, contrary to all other known interactions, resulting in a large number of free parameters. Besides determining quark masses, the interaction of the quarks with the Higgs is responsible for the peculiar pattern of mixing of the various families of quarks under weak interactions, and the corresponding hierarchy in the various decay modes of the heavier quarks into the lighter ones. In particular, the interplay of weak and Higgs interactions implies that processes with a change of flavour mediated by a neutral current (FCNC) can occur only at higher orders in the electroweak interactions and are strongly suppressed. This strong suppression make FCNC processes natural candidates to search for physics beyond the SM: if the new degrees of freedom do not have the same flavour structure of the quark-Higgs interaction present in the SM, then they could contribute to FCNC processes comparably to the SM amplitudes even if their masses are well above the electroweak scale, resulting in sizable deviations from the SM predictions for these rare processes. 

Observing new sources of flavour mixing is a natural expectation for most New Physics (NP) models.
While direct searches of new particles at high energies provide a direct information 
on the mass spectrum of the possible new degrees of freedom, the indirect information 
from low-energy flavour-changing processes translates into unique constraints on their couplings.  

During the first LHC run there has been a significant experimental progress in quark and lepton flavour physics. 
In the quark sector, the validity of the SM has been substantially
reinforced by a series of precision measurements in the $B_s$  and $B_d$ systems.
Altogether, the SM works remarkably well:  the Cabibbo-Kobayashi-Maskawa (CKM)~\cite{Cabibbo:1963yz,Kobayashi:1973fv} mechanism of quark-flavour 
mixing has been tested in various processes, although in many interesting cases the accuracy is still limited.
These results set stringent limits on the  flavour structure of physics beyond the SM,  and provide key
information for model-building. However, as we shall discuss in the following, several options are still open,  and 
the quality of this information could be substantially improved with refined studies of selected
flavour-violating observables (a complementary recent review on this subject can be found in Ref.~\cite{Buras:2013ooa}).

\section{The  flavour sector of the SM}
\label{sec:2}

The SM Lagrangian can be divided into two main parts, 
the gauge and the Higgs (or symmetry breaking) sectors. The gauge sector
is extremely simple and highly symmetric: it is completely specified by the 
local symmetry ${\mathcal G}^{\rm SM}_{\rm local} =SU(3)_{C}\times SU(2)_{L}\times U(1)_{Y}$
and by the fermion content,
\bea
\cL^{\rm SM}_{\rm gauge} &=& \sum_{i=1\ldots3}\ \sum_{\psi=Q^i_L \ldots E^i_R}
{\bar \psi} i \gamma^\mu D_\mu \psi  -\frac{1}{4} \sum_{a=1\ldots8} G^a_{\mu\nu} G^a_{\mu\nu} -\frac{1}{4}
 \sum_{a=1\ldots3} W^a_{\mu\nu} W^a_{\mu\nu}  -\frac{1}{4} B_{\mu\nu} B_{\mu\nu}~.
\eea
Here $G_{\mu\nu}^a$, $W_{\mu\nu}^a$, and $B_{\mu\nu}$ denote the 
field strength tensors of the three independent gauge groups in ${\mathcal G}^{\rm SM}_{\rm local}$, 
and $D_\mu$ the corresponding covariant derivative.
The fermion content consists of five fields with different quantum numbers 
under the gauge group,
\be
\label{eq:SMfer}
Q^i_{L}(3,2)_{+1/6}~,\ \ U^i_{R}(3,1)_{+2/3}~,\ \
D^i_{R}(3,1)_{-1/3}~,\ \ L^i_{L}(1,2)_{-1/2}~,\ \ E^i_{R}(1,1)_{-1}~,
\ee
each of them appearing in three different replica or 
flavours ($i=1,2,3$). The notation used above to indicate each field
is $\psi(A,B)_Y$, where $A$ and $B$ denote the representation under the 
$SU(3)_{C}$ and $SU(2)_L$ groups, respectively,  and $Y$ is the $U(1)_Y$ charge.

This structure gives rise to a large {\em global} flavour symmetry 
of $\cL^{\rm SM}_{\rm gauge}$.
Both the local and the global symmetries of $\cL^{\rm SM}_{\rm gauge}$
are broken with the introduction of a $SU(2)_L$ scalar doublet $\phi$,
or the Higgs field. The local symmetry is spontaneously 
broken by the vacuum expectation value of the Higgs field,
$\langle \phi \rangle  = v = (2\sqrt{2} G_F)^{-1/2} \approx 174$~GeV, where 
$G_F$ is the Fermi coupling. The global flavour symmetry is {\em explicitly broken} by 
the Yukawa interaction of $\phi$ with the fermion fields:
\be
\label{eq:SMY}
- {\cal L}^{\rm SM}_{\rm Yukawa}=Y_d^{ij} {\bar Q}^i_{L} \phi D^j_{R}
 +Y_u^{ij} {\bar Q}^i_{L} \tilde\phi U^j_{R} + Y_e^{ij} {\bar L}_{L}^i
\phi E_{R}^j + {\rm h.c.} \qquad  ( \tilde\phi=i\tau_2\phi^\dagger)~.  
\ee
The large global flavour symmetry of  $\cL^{\rm SM}_{\rm gauge}$, 
corresponding to the independent unitary rotations in flavour space 
of the five fermion fields in Eq.~(\ref{eq:SMfer}), is a $U(3)^5$ group. 
This can be decomposed as follows: 
\be 
{\mathcal G}_{\rm flavour} \equiv U(3)^5 = U(1)^5 \times  
{\mathcal G}_{q} \times {\mathcal G}_{\ell}~, 
\label{eq:Gtot}
\ee
where 
\be
{\mathcal G}_{q} = {SU}(3)_{Q_L}\times {SU}(3)_{U_R} \times {SU}(3)_{D_R}, \qquad 
{\mathcal G}_{\ell} =  {SU}(3)_{L_L} \otimes {SU}(3)_{E_R}~.
\label{eq:Ggroups}
\ee
Three of the five $U(1)$ subgroups can be identified with the total baryon and 
lepton numbers, which are not broken by $\cL_{\rm Yukawa}$, and the weak hypercharge, 
which is gauged and broken only spontaneously by $\langle \phi \rangle  
\not=0$. The subgroups controlling flavour-changing dynamics
and flavour non-universality  are the non-Abelian groups ${\mathcal G}_{q}$ 
and ${\mathcal G}_{\ell}$, which are explicitly broken by $Y_{d,u,e}$ not being 
proportional to the identity matrix. 

The diagonalization of each Yukawa coupling requires, in general, two 
independent unitary matrices, $V_L Y V^\dagger_R = {\rm diag}(y_1,y_2,y_3)$.
In the lepton sector the invariance of $\cL^{\rm SM}_{\rm gauge}$ 
under ${\mathcal G}_{\ell}$ allows us to freely choose the two matrices 
necessary to diagonalize  $Y_e$ without breaking gauge invariance, 
or without observable consequences. This is not the case in the quark 
sector, where we can freely choose only three of the four unitary matrices 
necessary to diagonalize both $Y_{d}$ and $Y_u$. Choosing the basis where 
$Y_{d}$ is diagonal (and eliminating the right-handed 
diagonalization matrix of $Y_u$) we can write 
\be
\label{eq:Ydbasis}
Y_d=\lambda_d~, \qquad  Y_u=V^\dagger\lambda_u~,
\ee
where 
\be
\label{eq:deflamd}
\lambda_d={\rm diag}(y_d,y_s,y_b)~, \ \ \
\lambda_u={\rm diag}(y_u,y_c,y_t)~, \qquad y_q = \frac{m_q}{v}~.
\ee
Alternatively we could choose a  gauge-invariant basis where 
$Y_d= V \lambda_d$ and $Y_u=\lambda_u$. Since the flavour symmetry  
does not allow the diagonalization from the left of both $Y_{d}$ and $Y_u$,
in both cases we are left with a non-trivial unitary mixing matrix, $V$, 
which is nothing but the CKM 
mixing matrix.

Eliminating unphysical phases related to quark-field redefinitions, it turns out that 
$V$ can be expressed in terms of four  physical parameters: three real angles and
one complex CP-violating phase. As a result, the full set of parameters controlling 
the breaking of the quark flavour symmetry in the SM is composed by the 
six quark masses in $\lambda_{u,d}$ and the four parameters of $V$.

For practical purposes it is often convenient to work in the mass eigenstate basis 
of both up- and  down-type quarks. This can be achieved rotating independently 
the up and down components of the quark doublet $Q_L$, or moving the CKM matrix 
from the Yukawa sector to the charged weak current in $\cL^{\rm SM}_{\rm gauge}$:
\be
\left. J_W^\mu \right|_{\rm quarks} = \bar u^i_L \gamma^\mu d^i_L \quad \stackrel{u,d~{\rm mass-basis}}{\longrightarrow} \quad
\bar u^i_L V_{ij} \gamma^\mu d^j_L ~.
\label{eq:Wcurrent}
\ee
However, it must be stressed that $V$ originates from the Yukawa sector (in particular 
from the misalignment of $Y_u$ and $Y_d$ in the ${SU}(3)_{Q_L}$ subgroup of ${\mathcal G}_q$): 
in the absence of Yukawa  couplings we can always set $V_{ij}=\delta_{ij}$. 

To summarize, 
quark flavour physics within the SM is characterized by a large flavour symmetry, 
${\mathcal G}_{q}$, defined by the gauge sector, whose only breaking sources 
are the two Yukawa couplings $Y_{d}$ and $Y_{u}$. The CKM matrix arises from the 
misalignment of $Y_u$ and $Y_d$ in flavour space.

\subsection{Some properties of the CKM matrix}

The standard parametrization of the CKM matrix~\cite{Chau:1984fp}
in terms of three rotational angles ($\theta_{ij}$) 
and one complex phase ($\delta$) is 
\bea
\label{eq:Chau}
V=\left(\begin{array}{ccc}
V_{ud}&V_{us}&V_{ub}\\
V_{cd}&V_{cs}&V_{cb}\\
V_{td}&V_{ts}&V_{tb}
\end{array}\right) 
=
\left(\begin{array}{ccc}
c_{12}c_{13}&s_{12}c_{13}&s_{13}e^{-i\delta}\\ -s_{12}c_{23}
-c_{12}s_{23}s_{13}e^{i\delta}&c_{12}c_{23}-s_{12}s_{23}s_{13}e^{i\delta}&
s_{23}c_{13}\\ s_{12}s_{23}-c_{12}c_{23}s_{13}e^{i\delta}&-s_{23}c_{12}
-s_{12}c_{23}s_{13}e^{i\delta}&c_{23}c_{13}
\end{array}\right)~,
\eea
where $c_{ij}=\cos\theta_{ij}$ and $s_{ij}=\sin\theta_{ij}$
($i,j=1,2,3$). 

The off-diagonal elements of the CKM matrix
show a strongly 
hierarchical pattern:  $|V_{us}|$ and $|V_{cd}|$ are close to $0.22$, the elements
$|V_{cb}|$ and $|V_{ts}|$ are of order $4\times 10^{-2}$ whereas $|V_{ub}|$ and
$|V_{td}|$ are of order $5\times 10^{-3}$. 
The Wolfenstein parametrization, namely the expansion of the CKM matrix 
elements in powers of the small parameter $\lambda \doteq |V_{us}| \approx 0.22$, is a
convenient way to exhibit this hierarchy in a more explicit way~\cite{Wolfenstein:1983yz}:
\begin{equation}
V=
\left(\begin{array}{ccc}
1-{\lambda^2\over 2}&\lambda&A\lambda^3(\varrho-i\eta)\\ -\lambda&
1-{\lambda^2\over 2}&A\lambda^2\\ A\lambda^3(1-\varrho-i\eta)&-A\lambda^2&
1\end{array}\right)
+O(\lambda^4)~,
\label{eq:Wolfpar} 
\end{equation}
where $A$, $\varrho$, and $\eta$ are free parameters of order 1. 
Because of the smallness of $\lambda$ and the fact that for each element 
the expansion parameter is actually $\lambda^2$, this is a rapidly converging
expansion.

The Wolfenstein parametrization is certainly more transparent than
the standard parametrization. If one requires sufficient 
level of accuracy, the terms of $O(\lambda^4)$ and 
$O(\lambda^5)$ have to be included in phenomenological applications.
This can be achieved in many different ways, according to the convention 
adopted. The simplest (and nowadays commonly adopted) choice is obtained by
{\it defining} the parameters $\{\lambda,A,\varrho,\eta\}$ in terms of 
the angles of the exact parametrization in Eq.~(\ref{eq:Chau}) as follows:
\begin{equation}
\label{eq:rhodef} 
\lambda\doteq s_{12}~,
\qquad
A \lambda^2\doteq s_{23}~,
\qquad
A \lambda^3 (\varrho-i \eta)\doteq s_{13} e^{-i\delta}~.
\end{equation}
The change of variables $\{ s_{ij}, \delta \} \to \{\lambda,A,\varrho,\eta\}$
in Eq.~(\ref{eq:Chau}) leads to an exact parametrization 
of the CKM matrix in terms of the Wolfenstein parameters. 
This parameterization can then be expanded to any given order
in powers of $\lambda$. In particular, one finds
\bea
V_{td} &=& A\lambda^3(1-\rhob-i\etab) +O(\lambda^7)~,  \no\\
V_{ts} &=&-A\lambda^2+\frac{1}{2}A\lambda^4[1-2 (\rhob+i\etab)]+ O(\lambda^6)~,  
\eea
where
\be
\rhob = \varrho (1-\frac{\lambda^2}{2})~,
\quad
\etab=\eta (1-\frac{\lambda^2}{2})~.
\ee

The unitarity of the CKM matrix implies the following relations between its
elements:
\be
{\bf I)}\quad 
 \sum_{k=1\ldots 3} V_{ik}^* V_{ki}=1~,
\quad\qquad 
{\bf II)}\quad
\sum_{k=1\ldots 3} V_{ik}^* V_{kj\not=i}=0~.
\ee
These relations are a distinctive feature of the SM, where the CKM matrix is the only 
source of quark flavour mixing.  Their experimental verification is therefore a useful 
tool to set bounds, or possibly reveal, new sources of flavour symmetry breaking. 
Among the relations of type {\bf II}, the one obtained for $i=1$ and $j=3$, 
namely 
\be
V_{ud}^{}V_{ub}^* + V_{cd}^{}V_{cb}^* + V_{td}^{}V_{tb}^* =0 
\label{eq:UT}
\ee
\be
{\rm or} \qquad 
\frac{V_{ud}^{}V_{ub}^*}{V_{cd}^{}V_{cb}^*}  + \frac{V_{td}^{}V_{tb}^*}{V_{cd}^{}V_{cb}^*}  + 1 = 0
\qquad \leftrightarrow \qquad
 -[\rhob+i \etab] - [(1-\rhob)-i\etab] +  1 =0~,
\no
\ee
is particularly interesting since it involves the sum of three terms all
of the same order in $\lambda$ and is usually represented as a unitarity triangle
in the complex  plane.
It is worth noting that Eq.~(\ref{eq:UT}) is invariant under any 
phase transformation of the quark fields. Under such transformations
the unitarity triangle is rotated in the complex plane,
but its angles and the sides remain unchanged.
Both angles and  sides of the unitary triangle are indeed observable quantities
which can be measured in suitable experiments.

\section{Flavour-changing processes occurring at the tree level}
\label{sec:4}

\begin{figure*}
\centering
\resizebox{0.90\textwidth}{!}{%
  \includegraphics{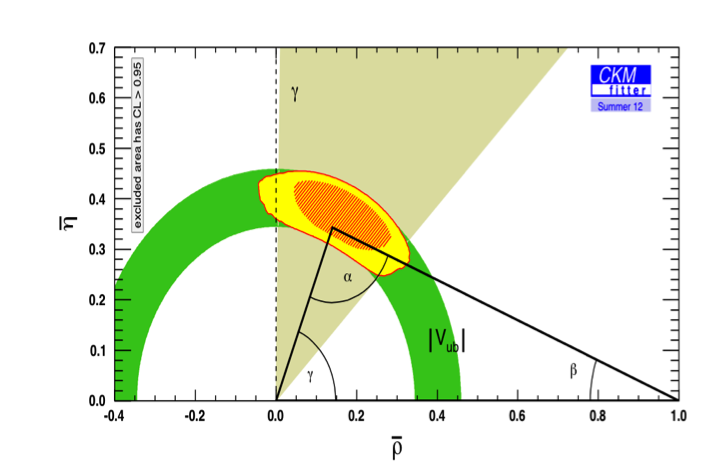}
  \includegraphics{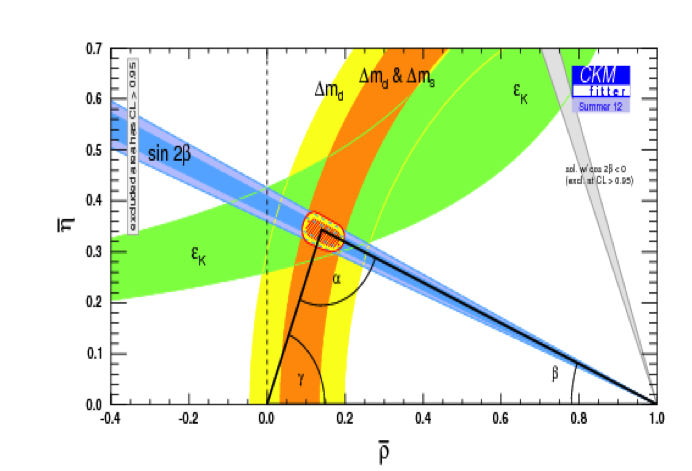}
}
\caption{Constraints on the ($\bar{\rho},\bar{\eta}$) plane obtained by the CKMfitter Collaboration~\cite{CKM_fitter}. 
The left panel shows the results obtained using only 
tree level measurements (Flavour Changing Charged Currents) and the right panel using the rest of loop level measurements (Flavour Changing Neutral Currents).}
\label{fig:4.1}
\end{figure*}

As discussed in Sect.~\ref{sec:2}, within the SM quarks are allowed to change flavour as a consequence 
of the Yukawa interaction. The couplings ruling this interaction, 
parameterized in terms of quark masses and CKM matrix elements,  are not predicted by the theory and need to be measured. 
If we assume that NP plays a relevant role only in processes occurring at the loop level within SM,  then the Yukawa couplings
(and in particular the CKM elements) can be  
determined to good accuracy by processes that occur at the tree level within the SM.
Constraints on NP are then obtained comparing the $\{\bar{\rho},\bar{\eta}\}$ values determined from processes 
dominated by tree-level diagrams (such as the measurement of $|V_{ub}|$ and its phase, discussed below) with the corresponding values 
determined by loop-induced amplitudes  (such as magnitude and phase of the $B_{d,s}$-meson mixing amplitude, discussed in the next section).

\subsection{$|V_{ub}/V_{cb}|$ }
\label{sec:4.1}
The ratio of the magnitudes of the CKM couplings $|V_{ub}|$/$|V_{cb}|$ has been measured at the B-factories 
using the flavour changing processes $b\rightarrow u(c) l \bar{\nu}$. The measured values of $|V_{ub}|$ using inclusive or exclusive 
methods show a discrepancy above two standard deviations, with the inclusive value being about $30\%$ larger~\cite{VubVcb}. Both methods suffer
from large theoretical and experimental uncertainties. The next generation 
B-factory experiment (Belle II) will produce hadronic tagged 
(i.e. using the fully hadronic decay mode of the other entangled $B_d$ decay in the event), 
high statistics and 
high purity samples which should allow to improve the situation. In the near future, 
LHCb is expected to provide competitive results in exclusive modes.

For some time the measured $BR(B^{\pm}\rightarrow\tau^{\pm}\nu)$, which can be seen as another 
determination of the ratio $|V_{ub}|$/$|V_{cb}|$, has been about three standard 
deviations higher than the one inferred from the CKM fit, in better agreement with the 
inclusive value of $|V_{ub}|$. However, in summer 2012, the Belle Collaboration updated 
their result with a much more precise hadron tag analysis , $BR(B^{\pm}\rightarrow\tau^{\pm}\nu) = (0.72^{+0.29}_{-0.27})\times10^{-4}$\cite{Btaunu_BELLE}. The world average becomes $BR(B^{\pm}\rightarrow\tau^{\pm}\nu) = (1.15\pm0.23)\times10^{-4}$ which is in better 
agreement with the fitted value $BR(B^{\pm}\rightarrow\tau^{\pm}\nu) = (0.74^{+0.09}_{-0.07})\times10^{-4}$~\cite{CKM_fitter}
(see also Ref.~\cite{UTfit}).
When this new result is incorporated into the fits, the overall consistency improves 
significantly and the value of the CKM parameters ($\bar{\rho}$ and $\bar{\eta}$) as determined 
by tree measurements gets closer to the values from loop measurements, see Fig.\ref{fig:4.1}.

On the other hand, the BaBar Collaboration in summer 2012 released a precise measurement 
of the ratio $BR(B_d\rightarrow D^{(*)}\tau\nu)$/$BR(B_d\rightarrow D^{(*)} l\nu)$~\cite{BDtaunu_BABAR} ($l=e,\mu$). 
The world average combination (dominated by the new BaBar result) is about three standard deviations 
higher than the result from the fit. There is no obvious NP explanation for this result compatible with other measurements.  
The Belle Collaboration should be able to provide a new result using a similar method that 
should clarify the situation.  

\subsection{The phase of $V_{ub}$: $\gamma$ }
\label{sec:4.2}
\begin{figure*}
\centering
\resizebox{0.90\textwidth}{!}{%
  \includegraphics{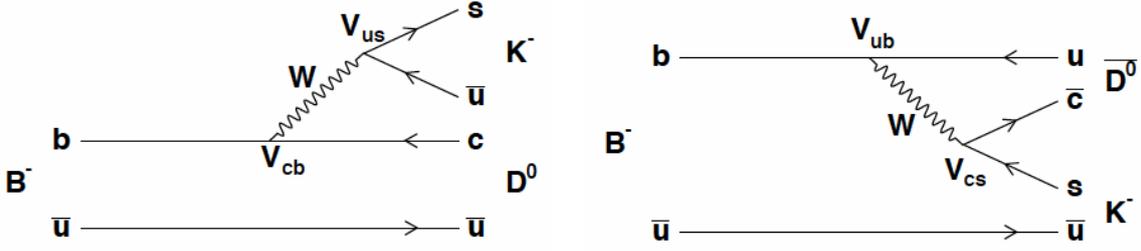}
}
\caption{Example of two $B$ decay amplitudes sensitive to the phase of $V_{ub}$ through their intereference when the $D$ and $\bar{D}$ mesons decay into the same final state.}\label{fig:4.2}
\end{figure*}

As can be seen in Fig.\ref{fig:4.1}, the determination of $\gamma$ is needed to determine the 
values of $\bar\rho$ and $\bar\eta$ at tree level. To a good approximation $\gamma$ at tree level can be 
determined as the phase in $b \rightarrow u$ transitions.  An example of these transitions is 
shown in Fig.\ref{fig:4.2}.  If the $D$ meson and the $\bar{D}$ meson decay into the same final state, 
the interference between the two amplitudes is sensitive to the phase of $V_{ub}$. In the example 
shown in Fig.\ref{fig:4.2} the experimental analysis is relatively simple, selecting and counting events 
to measure the ratio of $B^+$ and $B^-$ decays. However, the extraction of $\gamma$ from the measured 
ratio is a bit more involved as it requires the knowledge of the ratio of amplitudes for both the $B$ ($r_B$) and 
the $D$ decays ($r_D$), as well as the difference between the strong and weak phases involved ($\delta_{B}$ and $\delta_{D}$).
Several formalisms have been proposed which differ depending on the final state of the $D$ meson decay. If 
the $D$ meson decays into a CP eigenstate (for instance $K^+K^-$ or $\pi^+\pi^-$), then the relatively 
simple formalism denoted as "GLW" (Gronau, London and Wyler~\cite{GLW1,GLW2}) which does not depend on 
$\delta_{D}$ can be used. In the case of $D$ meson decays into non-CP eigenstates, like $K^+\pi^-$, the 
formalism is denoted as "ADS" (Atwood, Dunietz and Soni~\cite{ADS}) and requires external input from 
the charm factories to constrain $\delta_D$. A variation of this last method, denoted as "GGSZ" (Giri, Grossman, 
Soffer and Zupan~\cite{GGSZ}) exploits decays like $D\rightarrow K_s h^+ h^-$ to perform a Dalitz analysis 
and use the dependence of $\delta_D$ along the Dalitz phase space.

All these methods have been successfully used at the B-factories. The most precise determination of 
$\gamma$ is obtained from the Dalitz analysis of decays like $B^{\pm} \rightarrow D[K_s \pi^+ \pi^-] K^{\pm}$, 
i.e. using the "GGSZ" method. Combining all the different decay modes,
BABAR measures $\gamma = (69^{+17}_{-16})\,^{\circ}$~\cite{gamma_BABAR1,gamma_BABAR2} and 
Belle measures~$\gamma = (68^{+15}_{-14})\,^{\circ}$~\cite{gamma_BELLE1,gamma_BELLE2,gamma_BELLE3}. 
Recently, LHCb has released their 
first measurements of $\gamma$. In this case, the most sensitive analyses are the "GLW/ADS" due to the lower $K_s$ reconstruction efficiency. Their result, including a recent update of the "GGSZ" analysis with all the statistics on tape, can be quoted as $\gamma = (67 \pm 12)\,^{\circ}$~\cite{gamma_LHCb1,gamma_LHCb2,gamma_LHCb3}, in good agreement and with 
better precision than the results from the B-factories. However, as it should be clear from the discussion above, this is the result of a multiparameter fit and the comparison of the precision of just one of the parameters may be misleading due to the correlations, in particular with $r_B$. This measurement is also in good agreement with the determination of $\gamma$ from the fit using measurements at loop level: $\gamma = (66.6 \pm 6.4)\,^{\circ}$~\cite{CKM_fitter} (see also Ref.~\cite{UTfit}).

Most of the results from LHCb correspond to one third of their data already on tape and with only few decay modes analyzed. 
Therefore, there are high hopes to achieve a few degrees precision in the forthcoming years, which will allow for a 
meaningful comparison between loop and tree measurements. A good precision in both sides of Fig.\ref{fig:4.1} is 
required in order to exclude or confirm the presence of NP in loop processes.

\section{Flavour changing processes beyond the tree-level}
\label{sec:5}

As anticipated in Sect.~\ref{sec:1}, observing new sources of flavour mixing (i.e.~flavour violating couplings not related to quark and lepton
mass matrices) is a natural expectation for any extension of the SM
with new degrees of freedom not far from the TeV scale. It is also very natural to assume that these new sources of 
flavour mixing have a relatively larger impact in processes that are forbidden at the tree level within the SM. 
In the following our working hypothesis is that the measurements of $\bar{\rho}$ and $\bar{\eta}$ 
from tree-level processes are unaffected by NP, while NP effects can be sizable in amplitudes that are loop mediated 
within the SM. In this section, we will examine the experimental determination of loop mediated amplitudes, 
starting with measurements sensitive to the phase of the Yukawa couplings.
 
\subsection{CP Violation in B meson mixing}
\label{sec:5.1}

Loop processes denoted as "box" diagrams are responsible for the mixing phenomena in neutral mesons. The time evolution 
of a neutral $B$-meson is determined by a matrix with two components: the dispersive ($M$) and the absorptive ($\Gamma$) parts. 
The absorptive part is dominated by tree level diagrams with the production of real particles allowed by the momentum transfer ($q^2$) of the process. 
The dispersive part is sensitive to new particles contributing to the "box" diagram and it is where we may expect to see the effect of NP.

The frequency of the $B_{q}$ oscillations is determined by the magnitude of the off-diagonal elements of the dispersive matrix. The 
precise measurements of $\Delta m_{B_d} =  (0.510 \pm 0.004)~\mathrm{ps}^{-1}$~\cite{VubVcb} by the 
B-factories, and the recent precise determination of 
$\Delta m_{B_s}  =  (17.768 \pm 0.024)~\mathrm{ps}^{-1}$~\cite{DMS_LHCb} by 
the LHCb collaboration compared with the predictions from tree level measurements is a powerful constraint on the effect of NP in those loop diagrams, see Fig.\ref{fig:4.1}. The 
measurement constrains the combination $\bar{\rho}^2+\bar{\eta}^2$, although at this level of precision the dependence on $\bar{\eta}$ is very small.

While the magnitude of the couplings of new particles entering into the "box" diagrams is already severely constrained by the measurement 
of the oscillation frequency, it could still be a large effect from the phase of the couplings introduced by these new particles. A good way to 
access these potential new phases is through the measurement of CP asymmetries. The golden modes are $B_d\rightarrow J/\Psi K_s$ and 
$B_s\rightarrow J/\Psi \phi$ for the $B_d$ and $B_s$ systems respectively. Through the interference of the amplitudes where the original $B_{(d,s)}$ 
meson has or has not oscillated there is sensitivity to the phase of the $V_{td}$ ($V_{ts}$) coupling for the $B_d (B_s)$ systems respectively 
denoted as $\beta (-\phi_s/2)$ .

The final state $J/\Psi \phi$ can be a CP eigenstate with 
eigenvalue $\pm1$. The "plus" sign applies to an orbital angular momentum l=0 or 2, while the 'minus' sign is for l=1. 
In this case an angular analysis 
is needed to statistically disentangle the CP odd and even contributions. A tagging procedure is necessary to 
attribute each event to either $B_s$ or $\bar{B_s}$ decays.  The CP asymmetries measured at the B-factories in the $B_d$ system give, $\beta = (21.38^{+0.79}_{-0.77})\,^{\circ}$~\cite{sin2b_Bfactories1,sin2b_Bfactories2} which corresponds to a linear constraint in the 
$\bar{\rho}$ and $\bar{\eta}$ plane as can be seen in Fig.\ref{fig:4.1} consistent with the tree level measurements ($\beta = (24.9^{+0.8}_{-1.9})\,^{\circ}$~\cite{CKM_fitter}). However, the precision of the tree-level measurements does not allow to exclude NP phases contributing at the few degrees level in the $V_{td}$ coupling. 

\begin{figure*}
\centering
\resizebox{0.90\textwidth}{!}{%
 \includegraphics{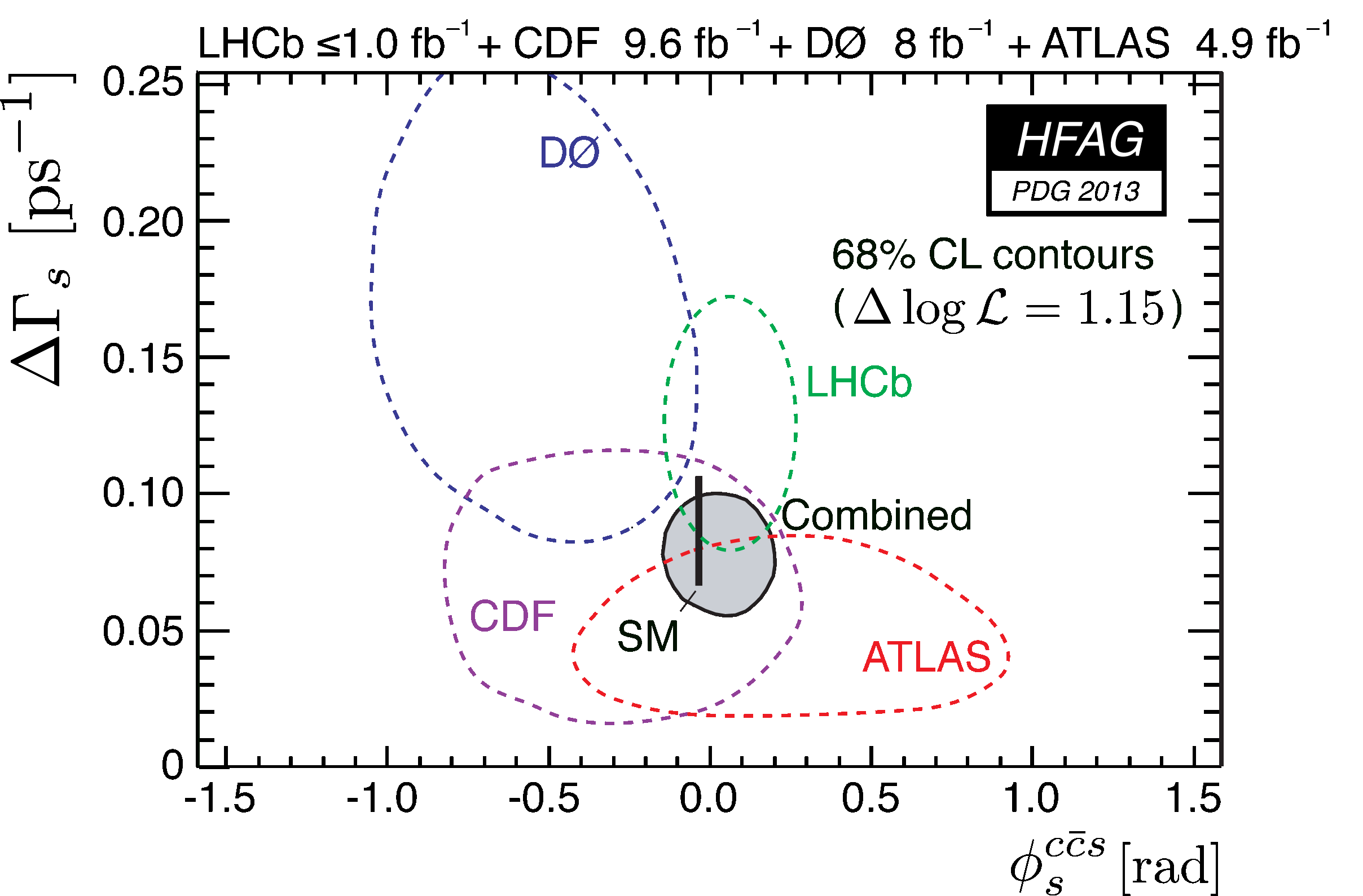}
  \includegraphics{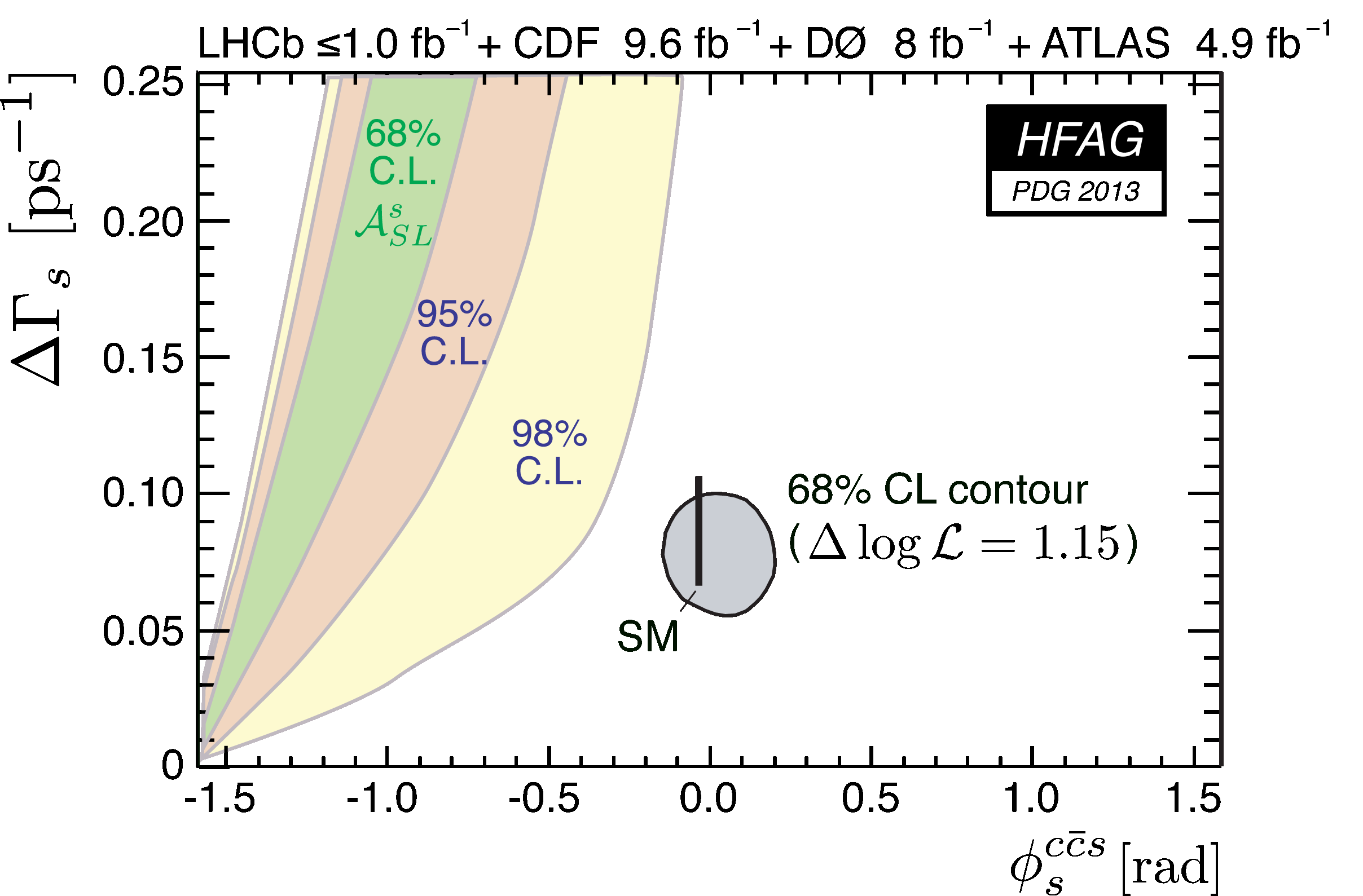}
}
\caption{The $68\%$ C.L. contours in the ($\phi_s$,$\Delta\Gamma_s$) plane, as obtained 
by the HFAG Collaboration~\cite{VubVcb}, where $\Delta\Gamma_s$ corresponds to the difference in the width between the two $B_s$ mass eigenstates and $\phi_s$ is defined in the main text. The left panel shows the 
individual contours of ATLAS, CDF, D0 and LHCb (in this figure the LHCb results using the $J/\Psi \phi$ final state corresponds to only 0.4/\fb), their combined contour (solid curve and shaded area), as well as the SM predictions. The right panel shows the same combined contour and SM predictions together with the regions allowed at $68\%$, $95\%$ and $98\%$ C.L. by the average measurements $A_{SL}(B_s) = -0.0171 \pm 0.0055$ and $\Delta m_{B_s}  = 17.69 \pm 0.08 \rm{ps}^{-1}$, through the relation tan$\phi_{12}$ = $A_{SL}(B_s) \times \Delta
m_{B_s} / \Delta\Gamma_s$, where $\phi_{12} =$ arg($-M_{12}/\Gamma_{12}$). This region is drawn under 
the assumption the phase difference ($\phi_s - \phi_{12}$) is equal to its SM prediction, see ~\cite{LenzNierste} for details.}
\label{fig:5.1}
\end{figure*}

Recently, the LHCb collaboration has precisely determined the analogue quantity in the $B_s$ system, $\phi_s = (0.6 \pm 4.0)\,^{\circ}$~\cite{sin2bs_LHCb1,sin2bs_LHCb2}, which can be compared with the determination from tree level measurements $\phi_s = (-2.3^{+0.1}_{-0.3}) \,^{\circ}$~\cite{CKM_fitter}. In this case, the precision on the tree level measurements is not the limiting factor and progress in the precision of the measurement, which is statistically dominated, is eagerly awaited. 
To a good approximation, $\phi_s$ is just a linear function of $\bar{\eta}$ therefore it is a direct constraint on the contributions of NP phases to the $V_{ts}$ coupling. On the left panel in Fig.\ref{fig:5.1} one can compare the measurements (dominated by the precision of the recent LHCb measurement) of $\phi_s$ with the expected value using the determination of $\bar{\eta}$ from the rest of flavour measurements (denoted as SM in Fig.\ref{fig:5.1}).

The measurements of CP asymmetries induced by the interference between the decay amplitudes and the mixing amplitudes in $B$-mesons constrain the effects of NP phases contributing to the box diagrams to few degrees.  However, few years ago D0 measured~\cite{D0_dimuon} an inclusive asymmetry defined as the difference between the number of $p\bar{p}$ collisions with two positive and two negative muons normalized to the total number of dimuon events. If the origin of the muons is assumed to be from semileptonic $B_{d,s}$-meson decays like $B _{d(s)}\rightarrow D_{(s)}^-\mu^+\nu $ and $\bar{B}_{d,s} \rightarrow B_{d,s} \rightarrow D_{(s)}^-\mu^+\nu$, this measurement is a linear combination of the "flavour-specific" asymmetries 
$a_{fs}$ for $B_d$ and $B_s$. This asymmetry is a direct measurement of CP violation in the mixing amplitude and in general can be written as:
$a_{fs} = |\Gamma_{12}/M_{12}|sin(\phi)$ where $\Gamma_{12}$ is the off-diagonal element of the absortive contribution and $M_{12}$ of the dispersive contribution. Within the SM, these asymmetries are very small for both $B_{d,s}$-mesons, and can be neglected within the experimental uncertainties. However, the D0 measurement of the inclusive asymmetry: $A^b_{SL} = -(0.787\pm0.172\pm0.093)\%$~\cite{D0_dimuon} deviates from zero by about $3.9\sigma$. It is relevant to notice that the systematic uncertainty quoted is relatively small because the background contribution to the dimuon asymmetry is estimated from the measurement of the equivalent single-muon asymmetry and the assumption that both background contributions are fully correlated. Other potential large systematic uncertainties like the detector asymmetry is controlled by the switch of the magnetic field polarity and production asymmetries are negligible due to the symmetry in the initial state: $p\bar{p}$ collisions.

On the other hand, the constraints from the previous measurements on the dispersive and absortive contributions leave very little space to accommodate the D0 measurement. Even considering that most of the effect could be NP contributing to the absorptive part in the $B_s$ system in $\Delta\Gamma_s$, the precision already achieved using $B_s\rightarrow J/\Psi \phi$ decays is not supporting such a scenario (see right panel of Fig.\ref{fig:5.1}). 
The B-factories have measured with relatively good precision 
$a_{fs}(B_d)=(-0.38\pm0.36)\%$~\cite{ASL_Bd_Bfactories1,ASL_Bd_Bfactories2}  which combined with other measurements of this quantity at D0 gives a world average of 
$a_{fs}(B_d)=(0.07\pm0.27)\%$~\cite{ASL_Bd_D0}  in good agreement with the SM expectations. Therefore, a large discrepancy is only possible in the $B_s$ system. 
The recent measurement from D0, 
$a_{fs}(B_s)=(-1.12\pm0.76)\%$~\cite{ASL_Bs_D0}, is not incompatible with the inclusive 
dimuon results neither with the more recent results from LHCb,  
$a_{fs}(B_s)=(-0.24\pm0.63)\%$~\cite{ASL_Bs_LHCb} giving a world average of 
$a_{fs}(B_s)=(-1.07\pm0.41)\%$ which is about $2.5\sigma$ away from zero (mainly driven by the D0 dimuon measurement).
LHCb needs to include more decay modes and more statistics to be able to conclude.
\label{sec:5.2}

\begin{figure*}
\centering
\resizebox{0.50\textwidth}{!}{%
\includegraphics{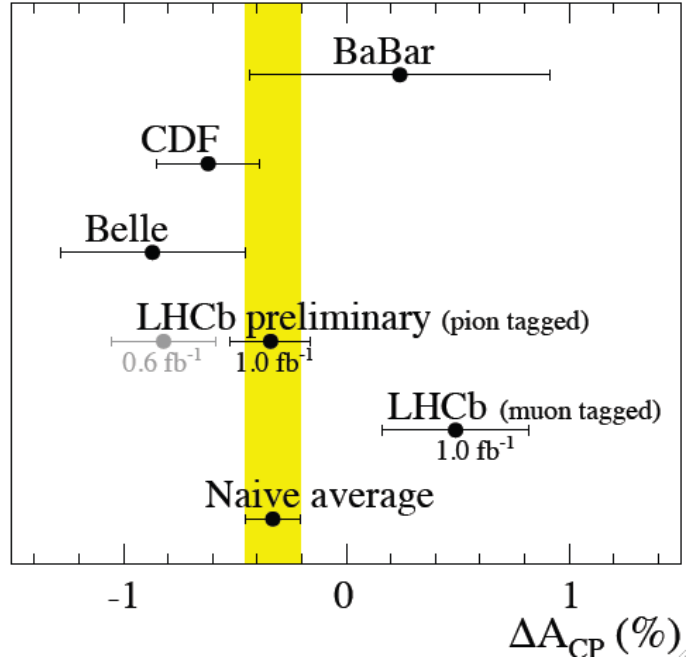}
}
\caption{Measurements of $\Delta A_{CP}$ as obtained by BABAR~\cite{BABAR_ACP}, CDF~\cite{CDF_ACP}, Belle~\cite{BELLE_ACP}, and LHCb~\cite{LHCb_ACP_0.6}~\cite{LHCb_ACP_1}~\cite{LHCb_ACP_semil}. The average has a probability of  $3.7\%$ to originate from statistical fluctuations.}\label{fig:5.2}
\end{figure*}

\subsection{CP violation in D meson decays}

Charm decays are complementary to $B$-meson decays as the particles contributing to the loops are down-type in the isospin doublets in contrast to the up-type in $B$-mesons. If NP differentiates between the up- and down-type quarks then it is interesting to compare $B$ and $D$ decays. 

So far there is no evidence for CP-violation in measurements of the CP asymmetries induced by the interference between mixing and decay amplitudes. However, the firsts LHCb  measurements of the direct CP asymmetries in $D\rightarrow \pi^+\pi^-$ and $D\rightarrow K^+ K^-$ decays indicated a somewhat surprising large effect. The LHCb measurement of 
$\Delta A_{CP} = A_{CP}(K^+K^-) - A_{CP}(\pi^+\pi^-)$ using self-tagged 
$D^{*\pm}\rightarrow D^0[h^+h^-]\pi^{\pm}$ decays indicated some evidence for a non-zero CP asymmetry: $\Delta A_{CP} = (-0.82 \pm 0.24)\%$~\cite{LHCb_ACP_0.6}. Later both CDF ($\Delta A_{CP} = (-0.62 \pm 0.23)\%$~\cite{CDF_ACP}) 
and Belle ($\Delta A_{CP} = (-0.87 \pm 0.41)\%$~\cite{BELLE_ACP}) 
confirmed this measurement and increased the hopes that NP could be finally seen at work.

The measurement of $\Delta A_{CP}$ is protected from large systematic uncertainties as to first order detector and production asymmetries cancel in the subtraction. Moreover, within the SM and the majority of NP models, there is no loss of sensitivity in the subtraction as the CP-asymmetry is expected to have opposite sign for $D\rightarrow K^+K^-$ and $D\rightarrow \pi^+ \pi^-$. Within the SM, the use of U-spin and QCD-factorization hypotheses leads to a prediction for $\Delta A_{CP}$ of about four times the ratio between the penguin amplitude (suppressed by $\lambda^5$) and the tree amplitude (only suppressed by $\lambda$), therefore the natural SM expectation is below $0.1\%$~\cite{Grossman:2006jg,Isidori:2011qw}.
NP effects could enlarge this asymmetry to the level seen experimentally, however the SM prediction also has sizable uncertainties 
(see e.g.~Ref.~\cite{Golden:1989qx, Brod:2011re,Pirtskhalava:2011va,mishima}). In particular,  
the U-spin approximation is already challenged by the fact that the branching ratio of $D\rightarrow \pi^+\pi^-$ is measured to be different (about three times larger) than the branching ratio of $D\rightarrow K^+K^-$. This fact could be the hint of large non-perturbative effects enhancing 
also the SM prediction of $\Delta A_{CP}$~\cite{Brod:2012ud}. As described in Ref.~\cite{Isidori:2012yx}, a valuable tool to distinguish SM vs.~NP  
contributions to $\Delta A_{CP}$ is provided, in principle, by the analysis of direct CP asymmetries in radiative modes
($D\rightarrow K^+K^-\gamma$ and $D\rightarrow \pi^+\pi^-\gamma$).

Recently, LHCb has updated the measurement including up to one third of the data already on tape. The new measurement, $\Delta A_{CP} = (-0.34 \pm 0.18)\%$~\cite{LHCb_ACP_1}, does not confirm the initial indications. Moreover, LHCb has performed a complementary analysis using self-tagged $B^\pm \rightarrow  D^0[h^+h^-]\mu^{\pm} \nu X$ events and measured $\Delta A_{CP} = (0.49 \pm 0.33)\%$~\cite{LHCb_ACP_semil}. The naive world average gives $\Delta A_{CP} = (-0.33 \pm 0.12)\%$ with an internal consistency that has a probability of $3.7\%$ to occur. The consistency of these measurements is ilustrated in Fig.\ref{fig:5.2}. The results from LHCb are completely dominated by the statistical uncertainty and therefore the situation will become more clear in 
the near future when the 3/\fb\ of data already on tape are analyzed. 

\subsection{Precise measurements in ElectroWeak (EW) penguin decays}
\label{sec:5.3}

\begin{figure*}
\centering
\resizebox{0.30\textwidth}{!}{%
 \includegraphics{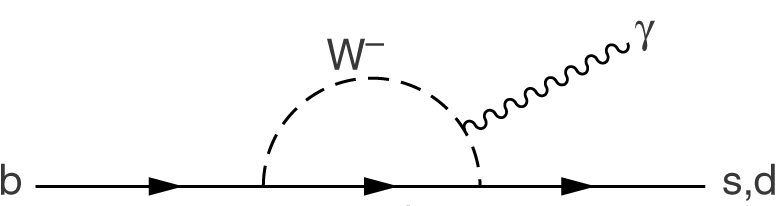}
}
\resizebox{0.30\textwidth}{!}{%
 \includegraphics{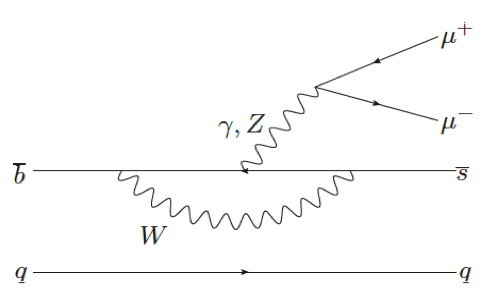}
}
\resizebox{0.30\textwidth}{!}{%
 \includegraphics{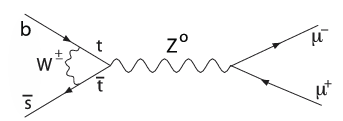}
}
\caption{Three realizations of the EW penguin. On the left panel the simplest form of radiative decays. On the center, the dominant contribution within the SM to the decay $B \rightarrow K^*\mu^+\mu^-$. On the right panel the dominant contribution within the SM to the very rare decay $B_s \rightarrow \mu^+ \mu^-$.}\label{fig:5.3}
\end{figure*}

The family of EW penguins involves only quarks, leptons and weak bosons to first order therefore the uncertainty due to QCD corrections is reduced compared with the dominant pure QCD penguins where the weak bosons are replaced by gluons. An example of its simplest realization can be seen in the left panel of Fig.~\ref{fig:5.3} and consist on a photon emited from the internal loop. If the photon decays into a lepton pair (hence the amplitude is further suppressed by a factor $\alpha_{\rm {QED}}$) or it is replaced by a Z boson, the process can provide a rich laboratory to test NP models. An example is shown on the center panel on Fig.~\ref{fig:5.3}.

The inclusive process $b\rightarrow s \gamma$ has been measured precisely at the 
B-factories, CESR and LEP with an uncertainty of $\sim 7\%$, $BR(b\rightarrow s\gamma) = (3.55 \pm 0.26)\times 10^{-4}$~\cite{PDG2012}, in agreement with the SM prediction,
$BR(b\rightarrow s\gamma) = (3.15 \pm 0.23)\times 10^{-4}$~\cite{bsgamma_SM}. In fact, this measurement is one of the strongest constraints in supersymmetric extensions of the SM.  Inclusive measurements are difficult at hadron colliders, but exclusive radiative decays are measured at LHCb with high statistics, (5300/\fb\ $B\rightarrow K^*\gamma$ and 
700/\fb\ $B_s\rightarrow \phi \gamma$ candidates). LHCb has already provided the most precise measurement of the radiative $B_s$ branching ratio: 
$BR(B_s\rightarrow \phi\gamma) = (3.5 \pm 0.4)\times 10^{-5}$~\cite{bsphigamma_LHCb}, however the theory predictions for these exclusive decays are not precise enough. In the near future, precise measurements of the photon polarization in these decays could be a promising search for NP effects.

\begin{figure*}
\centering
\resizebox{0.35\textwidth}{!}{%
 \includegraphics{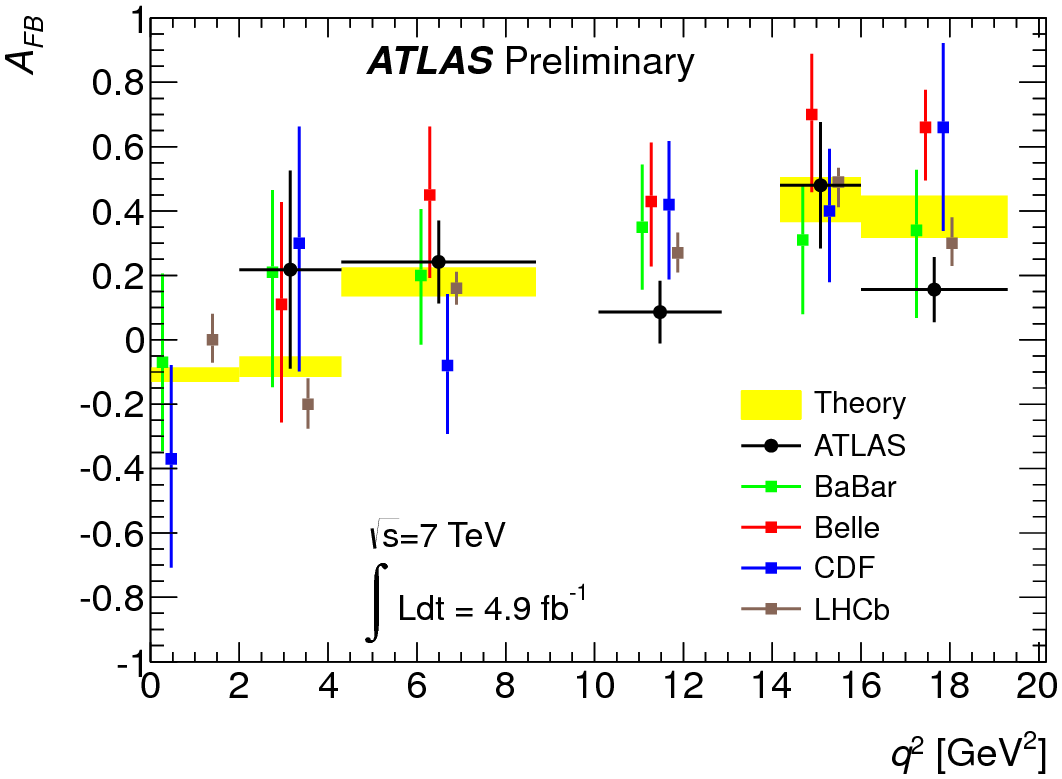}
}
\resizebox{0.40\textwidth}{!}{%
 \includegraphics{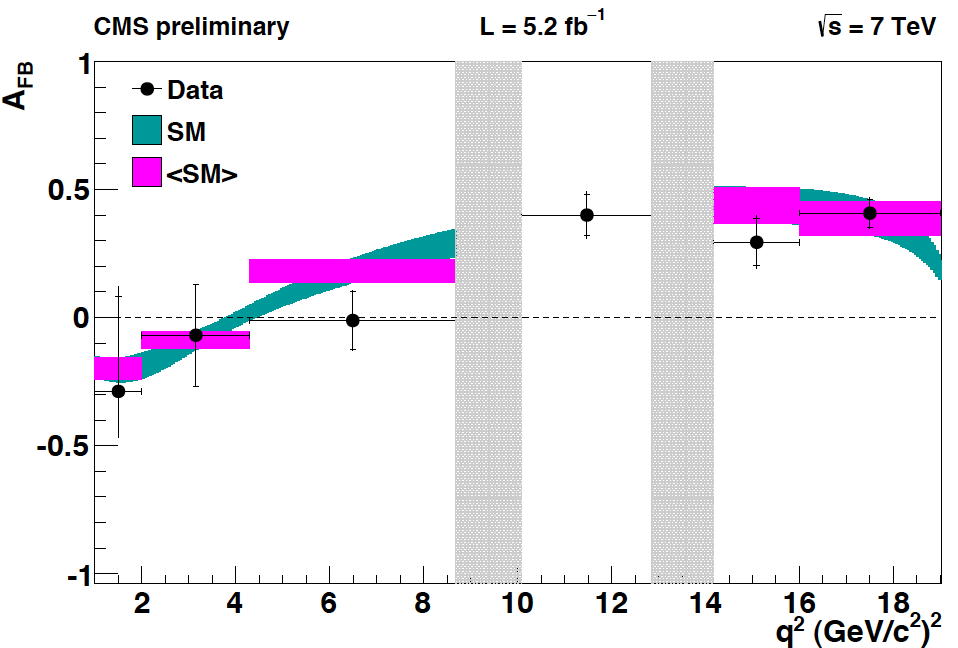}
}
\resizebox{0.45\textwidth}{!}{%
 \includegraphics{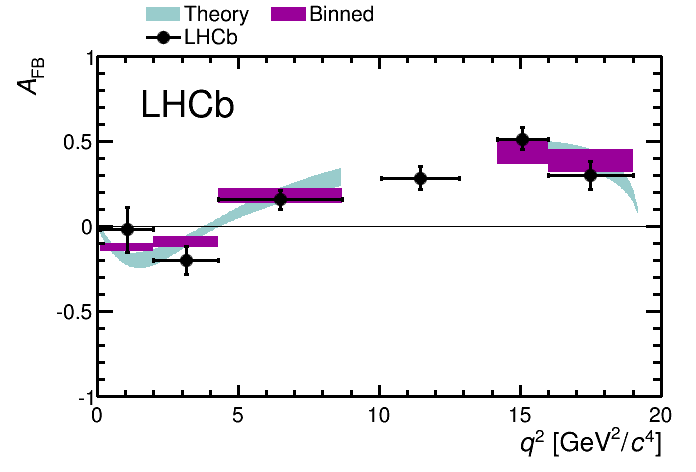}
}
\caption{Measurements of $A_{FB}$ as a function of $q^2$. On the top left panel, the ATLAS~\cite{Kmm_ATLAS} measurements are compared with previous measurements from B-factories, CDF and LHCb~\cite{Kmm_LHCb} (also shown in the bottom panel compared with the theoretical 
pre\-dictions~\cite{Bobeth:2011gi}). On the right top panel the CMS~\cite{Kmm_CMS} results are shown compared with the theoretical pre\-dictions from 
Ref.~\cite{Bobeth:2012vn}.}\label{fig:5.4}
\end{figure*}

The decay $B \rightarrow K^* \mu^+ \mu^-$ is the "golden mode" to test new vector(axial-vector) couplings contributing to the loops in $b \rightarrow s$ transitions, complementing the sensitivity to NP operators in radiative decays. The sign of the pion in the decay 
$K^* \rightarrow K^{\pm} \pi^{\mp}$ allows to tag the flavour of the $B$-meson, hence an angular analysis can be unambigously performed to test the helicity structure of the EW penguin. If we define $\theta_l$, as the polar angle of the negative lepton and the direction opposite that of the $\bar{B_d}$ in the dimuon rest frame,  $\theta_K$ the angle of the negative kaon in the $\bar{K^*}$ rest frame and $\phi$ the angle between the two planes defined by ($K,\pi$) and ($\mu^+,\mu^-$) in the $\bar{B_d}$ rest frame, we can write the differential branching ratio as a function of these angles and $q^2$ with only four parameters: $F_L, S_3, A_{FB}$ and $A_{IM}$. Hadronic uncertainties in the predictions for these parameters are 
under reasonable control. $F_L$ indicates the fraction of longitudinal polarization in $K^*$ and 
$A_{FB}$ measures the forward-backward asymmetry of the lepton. Both parameters are expected to have a strong dependence on $q^2$. $S_3$ measures the asymmetry in the 
$K^*$ transverse polarization and $A_{IM}$ is proportional to the T-odd CP asymmetry. 
These last two parameters are expected to be very small and mostly independent of $q^2$ given the current level of precision. The first angular analyses of the decay  $B \rightarrow K^* \mu^+ \mu^-$ performed at the BaBar, Belle and CDF experiments were limited by the large statistical uncertainties~\cite{Kmm_old1,Kmm_old2,Kmm_old3}. Recently, the LHC experiments have taken over.  The LHCb experiment~\cite{Kmm_LHCb} with only 1~\fb\ has the largest sample of $B \rightarrow K^* \mu^+ \mu^-$ candidates (900/\fb) with very small background and therefore has the best sensitivity, but also ATLAS~\cite{Kmm_ATLAS} and CMS~\cite{Kmm_CMS} with $\sim$80/\fb\ candidates are competitive, in particular at large values of $q^2$. In Fig.~\ref{fig:5.4} the measurements of the parameter $A_{FB}$ by the LHC experiments are compared with previous measurements and the SM predictions. The precision achieved by the LHCb experiment allows for a determination of the zero crossing-point for the first time: $q^2(A_{FB}=0) = 4.9 \pm 0.9~\rm{GeV}^2$ which is in good agreement with the relatively more precise SM prediction: $4.4\pm0.3~\rm{GeV}^2$~\cite{SM_AFB0}.\footnote{~Kinematical studies of  $B \rightarrow K^* \mu^+ \mu^-$ at LHCb have significantly improved 
after the completion of this review, that reflects the status of flavor-physics observables before 
the summer 2013. In particular, a significant discrepancy between SM predictions and the so-called 
$P_5^\prime$ asymmetry~\cite{Descotes-Genon:2013vna}
in the low-$q^2$ region has recently been observed at LHCb~\cite{Aaij:2013qta}.
This result has triggered a significant amount of theoretical/phenomenological work, 
mainly focused on a reliable re-analysis of the (SM) theoretical error 
on $P_5$~\cite{Altmannshofer:2013foa,Hambrock:2013zya,Descotes-Genon:2013zva,Hurth:2013ssa,Beaujean:2013soa}, that is likely to 
be underestimated (see also Ref.~\cite{Jager:2012uw}). We refer to these recent papers for a more detailed discussion about this point.}

There are other observables and many other decay modes sensitive to NP affecting the EW penguin which are being studied or will be available with more statistics. The increase in 
precision of these analyses will allow for one of the strongest tests of generic NP models, up to some level independent of the NP flavour structure.

\subsection{Pure leptonic penguin decays}
\label{sec:5.4}

\begin{figure*}
\centering
\resizebox{0.4\textwidth}{!}{%
 \includegraphics{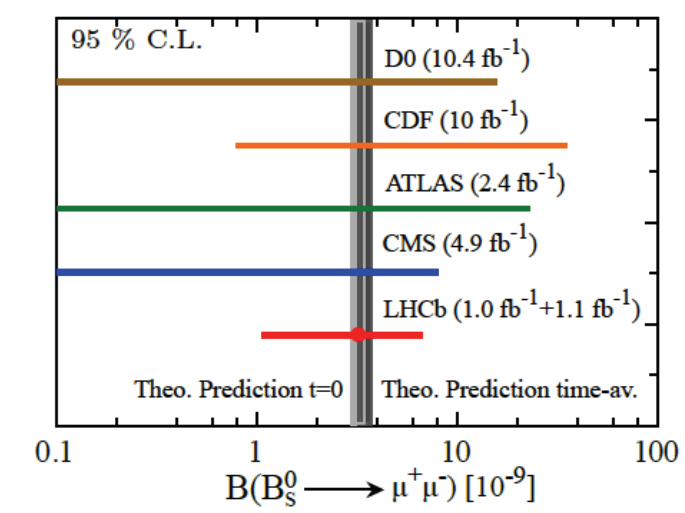}
}
\resizebox{0.45\textwidth}{!}{%
\includegraphics{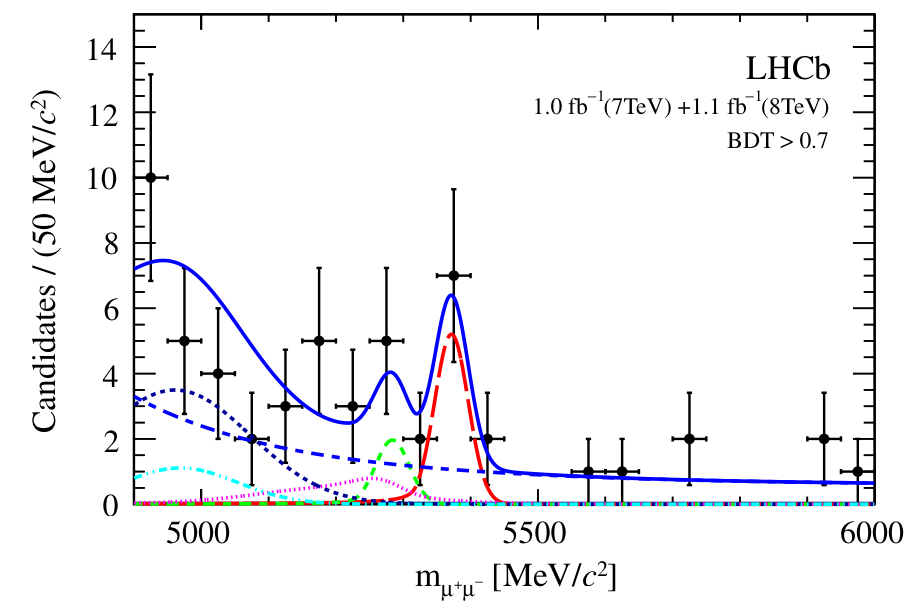}
}
\caption{On the left panel the $95\%$ C.L. intervals as determined by D0, CDF, ATLAS, CMS and LHCb. On the right panel the invariant mass distribution for events with a high value of the multivariate discriminant (BDT$>$0.7). The result of the fit is overlaid (blue solid line) and the different components detailed: $B_s \rightarrow \mu^+ \mu^-$ (red long dashed curve), $B_d \rightarrow \mu^+ \mu^-$ (green medium dashed curve), $B_{(s)} \rightarrow h^+ h^{,-}$ (pink dotted curve), $B_d \rightarrow \pi^- \mu^+ \nu$ (black short dashed curve), and $B^{(+)} \rightarrow \pi^{(+)} \mu^+ \mu^-$ (light blue dash-dotted curve), and the combinatorial background (blue medium dashed).}\label{fig:5.5}
\end{figure*}

The pure leptonic decays of $K, D$ and $B$-mesons are a particularly interesting case 
of EW penguins, see right panel of Fig.~\ref{fig:5.3}. Compared with the decays 
described in section~\ref{sec:5.3}, the 
helicity configuration of the final state suppresses the vector(axial-vector) contribution by 
a factor proportional to $({m_l} / {M_{K,D,B}})^2$. Therefore, these decays are 
particularly sensitive to new (pseudo-)scalar interactions. This is why they are 
sometimes referred as "Higgs-penguins"~\cite{Higgspenguin}.

In the case of $K$ and $D$-meson decays, the contribution of the absorptive part in 
the amplitude can be dominant. Indeed, this is what is observed in the measurement 
of $BR(K_L \rightarrow \mu^+\mu^-)=(6.84\pm0.11)\times 10^{-9}$~\cite{PDG2012}. The equivalent contribution to 
the $K_s$ decay is calculated to be $\sim 5\times10^{-12}$~\cite{Ksmm_SM1,Ksmm_SM2}, therefore if 
$BR(K_s \rightarrow \mu^+\mu^-)$ is measured to be larger than $10^{-11}$ it should be a clear indication of NP in the dispersive contribution, and should certainly increase the interest in measurements like the $BR(K^+\rightarrow \pi^+ \nu \nu)$. 
The excellent invariant mass resolution of the LHCb experiment, together with the large production of kaons at the LHC ($10^{13} K_s$/\fb) has allowed LHCb to improve drastically (by a factor 30) the limit on $BR(K_s \rightarrow \mu^+ \mu^-)$. With 1~\fb\ of data LHCb obtains a limit of  $BR(K_s \rightarrow \mu^+ \mu^-)< 9\times 10^{-9}$ at 
$90\%$CL~\cite{Ksmm_LHCb}, similar to the limit obtained by KLOE on 
$BR(K_s \rightarrow e^+ e^-) < 9\times 10^{-9}$ at $90\%$CL~\cite{Ksee_KLOE}.
 
In the case of D-meson decays the absorptive contribution is limited by the measured 
$BR(D \rightarrow \gamma \gamma)$ at BABAR~\cite{D2gg_BABAR} to be up to $6\times10^{-11}$~\cite{D2mm_absorptive}. Again the excellent LHCb detector performance 
and large charm production at the LHC has allowed to improve the existing results (by a factor 20), $BR(D \rightarrow \mu^+ \mu^-) < 6.2\times 10^{-9}$ at 
$90\%$CL~\cite{D2mm_LHCb}. Although  these are certainly large improvements, the LHCb results have been obtained with only one third of the data already available. The experiment expects to collect the equivalent of two orders of magnitude more data in the next decade (taking into account the increase in cross-section at higher energies). Therefore, we should expect that the interesting region between 
$10^{-11}$ and $10^{-9}$ is explored in the next decade for $K$- and $D$-mesons decays.

In the case of $B_d$ and $B_s$-meson decays the contribution of the absorptive part can be safely neglected. These decays are well predicted theoretically and experimentally are exceptionally clean, in particular the $B_s$ decay. Using tree level measurements as described in Sect.~\ref{sec:4}, the SM predictions are~\cite{Bobeth:2013uxa1,Bobeth:2013uxa2,Bobeth:2013uxa3} (see also~~\cite{Buras:2012ru}):
$BR(B_s \rightarrow \mu^+ \mu^-) = (3.65\pm0.23)\times 10^{-9}$ and 
$BR(B_s \rightarrow \mu^+ \mu^-) = (1.06\pm0.09) \times 10^{-10}$. In the $B_s$ case, this prediction corresponds to a flavour-averaged 
time-integrated measurement, taking into account the 
correction due to the non-vanishing width difference between $B_s$ and $\bar B_s$ mesons~\cite{Bsmm_DGscorrection}.  As mentioned before, these decays are superb tests for new (pseudo-)scalar 
contributions. In particular, within the Minimal Supersymmetric Standard Model the purely leptonic $B$-meson branching ratios are proportional to $\tan^6\beta/M^4_A$,  where $\tan\beta$ denotes the ratio of the two Higgs vacuum expectation values, and $M_A$ the mass of the pseudo-scalar Higgs. Therefore these decays are very sensitive probes of the large "$\tan \beta$" regime (see Sect.~\ref{sec:6} for more details). 

The main difficulty of the experimental analysis is the very large ratio between the background and the expected signal. For instance, if we assume the SM value branching ratio after the trigger and selection procedures, CDF~\cite{Bsmm_CDF} expects 
$\sim 0.26~B_s \rightarrow \mu^+ \mu^-$ candidates per \fb, ATLAS~\cite{Bsmm_ATLAS} $\sim 0.4$, CMS~\cite{Bsmm_CMS} $\sim 0.8$ and LHCb~\cite{Bsmm_LHCb} $\sim 3.6$ (with the multivariate discriminant (BDT)$>$0.7). The background (in the $B_s$ mass window) is dominated by combinations of real muons and it is estimated from the mass sidebands. Therefore the main handle to reduce the combinatorial background is the invariant mass resolution of the experiments, where a factor two better resolution is equivalent to a factor two more luminosity: ATLAS 80~MeV/$\rm{c}^2$, CMS 45~MeV/$\rm{c}^2$, CDF 25~MeV/$\rm{c}^2$  and LHCb 22~MeV/$\rm{c}^2$. Taking into account the previous estimates 1~\fb\ at LHCb is roughly equivalent to 10~\fb\ at CMS and 
20~\fb\ at ATLAS/CDF for the same analysis strategies. LHCb is also using the shape of the signal probability in invariant mass and multivariate discriminant distributions from control channels ($B_d \rightarrow h^+h^{'-}$), rather than doing a simple counting experiment. 

The results of the search for the decay $B_s \rightarrow \mu^+ \mu^-$ from the Tevatron~\cite{Bsmm_CDF,Bsmm_D0} 
and LHC~\cite{Bsmm_ATLAS,Bsmm_CMS,Bsmm_LHCb} 
experiments is summarized at the left panel of Fig.~\ref{fig:5.5}. CDF observes a 
slight excess with respect to the background only hypothesis with a probability of $\sim 0.9\%$ while the LHCb sensitivity allows to quote a first evidence for this decay with the probability of being a background fluctuation of $\sim 0.0005\%$ (equivalent to $3.5\sigma$). LHCb determines 
$BR(B_s \rightarrow \mu^+ \mu^-) = (3.2^{+1.5}_{-1.2}) \times 10^{-9}$ in excellent agreement with the SM expectation. The invariant mass distribution obtained by 
LHCb can be seen at the right panel in Fig.~\ref{fig:5.5}. The search for the decay 
$B_d \rightarrow \mu^+ \mu^-$ does not show a significant excess with respect to the background only hypothesis, and the best limit is obtained by LHCb to be:
$BR(B_d \rightarrow \mu^+ \mu^-) < 9.4 \times 10^{-10}$ at $95\%$CL.

Very recently the CMS and LHCb Collaborations have updated their results for the summer 2013 conferences including the full datasets ($\sim25$~\fb\ from CMS and $\sim3$~\fb\ from LHCb). Both collaborations see clear evidence for the 
decay $B_s \rightarrow \mu^+ \mu^-$ with significances larger or equal to $4\sigma$ for each individual measurement. The CMS measurement~\cite{Bsmm_CMS_new}, 
$BR(B_s \rightarrow \mu^+ \mu^-) = (3.0^{+1.0}_{-0.9}) \times 10^{-9}$ and the LHCb measurement~\cite{Bsmm_LHCb_new}, $BR(B_s \rightarrow \mu^+ \mu^-) = (2.9^{+1.1}_{-1.0}) \times 10^{-9}$ are compatible and also in agreement with the SM.
A preliminary combination of these two measurements gives~\cite{Bsmm_combination},
$BR(B_s \rightarrow \mu^+ \mu^-) = (2.9\pm0.7) \times 10^{-9}$ which corresponds to a clear observation of this decay in agreement with the SM expectation. In the case of the $B_d$ decays, both collaborations do not claim yet a significant excess and quote a limit, $BR(B_d \rightarrow \mu^+ \mu^-) < 1.1 \times 10^{-9}$ at $95\%$CL from CMS and 
$BR(B_d \rightarrow \mu^+ \mu^-) < 7.4 \times 10^{-10}$ at $95\%$CL from LHCb.

\section{Implications for New Physics models}
\label{sec:6}

\subsection{General considerations}

If physics beyond the SM respects the SM gauge symmetry, as we expect from general 
arguments, the low-energy 
amplitudes  describing the transition of a fermion $\psi_i$ to a fermion $\psi_j$ (of different flavour)
can be decomposed in the following general form
\begin{equation}
\mathcal{A}(\psi_i \to \psi_j + X) = \mathcal{A}_{0} \left[  \frac{c_{\rm SM}}{M_W^2} 
+ \frac{c_{\rm NP}}{\Lambda^2} \right]~,
\label{eq:fl1}
\end{equation} 
where $\Lambda$ is the energy scale of the new degrees of freedom and 
the SM result is recovered in the limit $c_{\rm NP}\to 0$.
This structure is completely general: the coefficients 
$c_{\rm SM(NP)}$ may include appropriate CKM coefficient factors 
and eventually a $\sim 1/(16\pi^2)$ suppression 
if the amplitude is loop-mediated. Given our ignorance 
about the $c_{\rm NP}$ coefficients, the values of the scale $\Lambda$ probed by present 
experiments vary over a wide range. However, the general result 
in Eq.~(\ref{eq:fl1}) allows us to predict how these bounds will 
improve with future experiments: increasing the statistic on a given observable, 
the corresponding bound on  $\Lambda$ 
scales at most as $N^{1/4}$, where $N$ is the relative increase in the number of events 
used to measure the observable.\footnote{Let's consider 
separately forbidden SM processes ($c_{\rm NP} \gg c_{\rm SM}$)  and processes where 
NP is expected to be a small correction over the SM ($c_{\rm NP} \ll c_{\rm SM}$). 
In the first case, assuming negligible background, 
the experimental bound on the forbidden rate 
scales linearly with $N$ and the predicted rate of the process scales as $1/\Lambda^4$.
In the second case, assuming a positive signal has already been observed, 
the improvement on the experimental precision scales as $N^{-1/2}$ and the bound is set on the interference term 
between SM and NP, that scales as $1/\Lambda^2$. As a result, in both cases the sensitivity on $\Lambda$ 
scales as $N^{1/4}$.}
From  Eq.~(\ref{eq:fl1}) it is also clear that 
indirect searches can probe  NP scales well above the TeV
for models where ($c_{\rm SM} \ll c_{\rm NP}$),
namely models which do not respect the symmetries and 
the symmetry-breaking pattern of the SM.

\begin{figure*}
\centering
\resizebox{0.45\textwidth}{!}{%
 \includegraphics{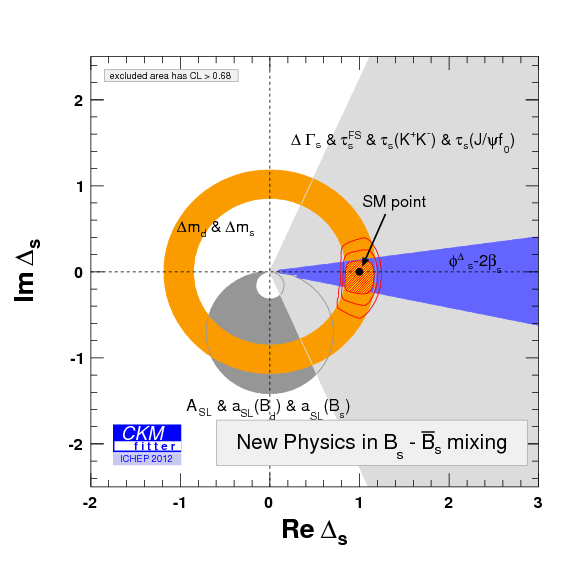}
}
\resizebox{0.45\textwidth}{!}{%
 \includegraphics{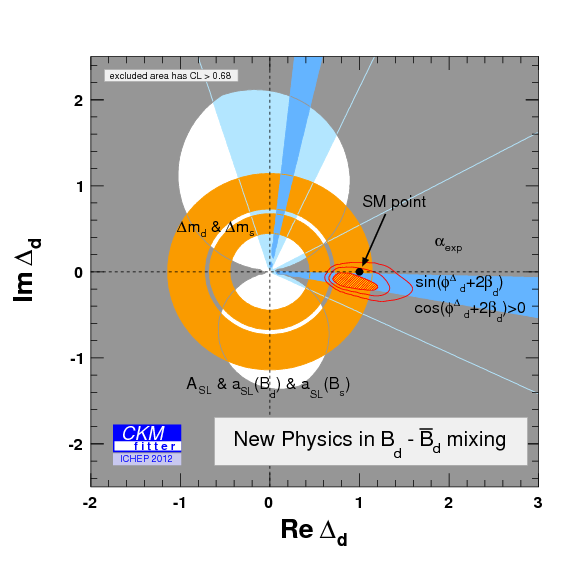}
}
\caption{ Model independent fit in the scenario where NP affects $\bar B_d$--$B_d$ 
and $\bar B_s$--$B_s$ mixing amplitudes separately~\cite{Lenz:2012az}. The
coloured areas represent regions with C.L.~$< 68.3 \%$ for the individual constraints. 
The red area shows the region with C.L.~$< 68.3 \%$ for the combined fit, 
with the two additional contours delimiting the regions with C.L.~$< 95.45 \%$ and C.L.~$< 99.73 \%$
(see Ref.~\cite{Lenz:2012az} for more details).}
 \label{fig:bcpv:newphysics}
\end{figure*}

As discussed in the previous section, the main strategy to search (or to constrain) physics beyond the SM in
flavour-violating observables consists of two steps: i)~determine the CKM elements from processes that are 
tree-level dominated within the SM (such that ${c_{\rm SM}} \gg {c_{\rm NP}}$), ii)~use these values to predict 
loop-mediated amplitudes and compare them with the corresponding measurements, allowing for non-vanishing NP
effects. An illustration of this procedure in the case of  the two $B_{d,s}$-meson mixing 
amplitudes is shown in Fig.~\ref{fig:bcpv:newphysics}~\cite{Lenz:2012az}. The possible NP contributions to $B_{s,d}$-meson mixing 
are parameterized in terms of two complex parameters, 
 $\Delta_{s,d}$,   describing the normalization of the amplitude with respect to the SM case (the SM is recovered for $\Delta_s=\Delta_d=1$).
The results of the fit to data allowing for generic  $\Delta_{s,d}$, is reported in Fig.~\ref{fig:bcpv:newphysics}. 
As can be seen, the present data are in good agreement with the SM expectation. On the other hand, 
non-vanishing NP contributions  of $O(20\%)$ for both $|\Delta_s|$ and $|\Delta_d|$, and up to $0.1$ and 
$0.05$ for $\phi_s={\rm arg}(\Delta_s)$  and  $\phi_d={\rm arg}(\Delta_d)$, respectively,  
are still allowed.

Assuming NP is heavy, the experimental bounds on the effective couplings $\Delta_{s,d}$ can be translated into bounds on the couplings 
of effective NP local operators (generated by the exchange of the heavy new states) 
contributing to the $B_{s,d}$-meson mixing amplitudes. In     
Table~\ref{tab:DF2} we report such bounds for the most representative $\Delta F=2$ operators (i.e.~operators 
contributing to meson-antimeson mixing at the tree level), together with similar bounds obtained from meson-antimeson 
mixing in the kaon and $D$-meson systems.\footnote{For the definition of the observables describing CP violation in the neutral kaon 
system ($\epsilon_K$ and $\epsilon^\prime$) and those describing CP violation in the $D$-meson mixing amplitude 
($|q/p|_D$ and $\phi_D$) we refer to Ref.~\cite{D'Ambrosio:1996nm}
and \cite{Gedalia:2009kh}, respectively. Updated constraints on  $D$-meson mixing parameters can be found in 
Ref.~\cite{oai:arXiv.org:1206.6245}.}
As can be seen, for $c_{\rm NP}=1$   present data probes very high scales.
On the other hand, if we insist with the theoretical prejudice that NP must show up 
not far from the TeV scale in order to stabilize the Higgs sector, then the new degrees of freedom 
must have a  peculiar flavour structure able to justify the smallness of the effective couplings 
$c_{\rm NP}$ for $\Lambda=1$~TeV (this fact is often known as the NP {\em flavour problem}).

\begin{table}[t]
\begin{center}
\begin{tabular}{c|c c|c c|l} \hline\hline
\rule{0pt}{1.2em}%
Operator &  \multicolumn{2}{c|}{Bounds on $\Lambda$~in~TeV~($c_{\rm NP}=1$)} &
\multicolumn{2}{c|}{Bounds on
$c_{\rm NP}$~($\Lambda=1$~TeV) }& Observables\cr
&   Re& Im & Re & Im \cr  
 \hline $(\bar s_L \gamma^\mu d_L )^2$  &~$9.8 \times 10^{2}$& $1.6 \times 10^{4}$ 
&$9.0 \times 10^{-7}$& $3.4 \times 10^{-9}$ &  \raisebox{-5pt}[0pt][0pt]{ $\Delta m_K$; \ \ $\epsilon_K$} \\ 
($\bar s_R\, d_L)(\bar s_L d_R$)   & $1.8 \times 10^{4}$& $3.2 \times 10^{5}$ 
&$6.9 \times 10^{-9}$& $2.6 \times 10^{-11}$ &  \\ 
 \hline $(\bar c_L \gamma^\mu u_L )^2$  &$1.2 \times 10^{3}$& $2.9 \times 10^{3}$ 
&$5.6 \times 10^{-7}$& $1.0 \times 10^{-7}$ &  \raisebox{-5pt}[0pt][0pt]{ $\Delta m_D$; \ \ $|q/p|_D$, $\phi_D$} \\ 
($\bar c_R\, u_L)(\bar c_L u_R$)   & $6.2 \times 10^{3}$& $1.5 \times 10^{4}$ 
&$5.7 \times 10^{-8}$& $1.1 \times 10^{-8}$ &  \\ 
\hline$(\bar b_L \gamma^\mu d_L )^2$    &  $6.6 \times 10^{2}$ & $ 9.3 \times 10^{2}$ 
&  $2.3 \times 10^{-6}$ &
$1.1 \times 10^{-6}$ & \raisebox{-5pt}[0pt][0pt]{  $\Delta m_{B_d}$;\ \  $\sin(2\beta)$ from $B_d \to \psi K$}  \\ 
($\bar b_R\, d_L)(\bar b_L d_R)$  &   $  2.5 \times 10^{3}$ & $ 3.6
\times 10^{3}$ &  $ 3.9 \times 10^{-7}$ &   $ 1.9 \times 10^{-7}$ 
&    \\
\hline $(\bar b_L \gamma^\mu s_L )^2$    &  $1.4 \times 10^{2}$ &  $  2.5 \times 10^{2}$   &  
 $5.0 \times 10^{-5}$ &   $1.7 \times 10^{-5}$ 
   &  \raisebox{-5pt}[0pt][0pt]{  $\Delta m_{B_s}$;\ \  $\sin(\phi_s)$ from $B_s \to \psi \phi$} \\ 
($\bar b_R \,s_L)(\bar b_L s_R)$  &    $ 4.8  \times 10^{2}$ &  $ 8.3  \times 10^{2}$  & 
   $8.8 \times 10^{-6}$ &   $2.9 \times 10^{-6}$  
  & \\ \hline\hline
\end{tabular}
\caption{\label{tab:DF2} Bounds on representative dimension-six  $\Delta F=2$  
operators~\cite{Isidori:2010kg,Isidori:2013ez}. The bounds on $\Lambda$ are evaluated 
assuming an effective coupling $1/\Lambda^2$ (i.e.~setting $c_{\rm NP}=1$).
Alternatively, the bounds on the respective $c_{\rm NP}$ are obtained assuming
$\Lambda=1$ TeV. In the last column we list the observables used to set such bounds; 
the observables related to CPV are 
separated from the CP conserving ones with semicolons.  }
\end{center}
\end{table}

\subsection{Minimal Flavour Violation}

A natural suppression of the $c_{\rm NP}$ is obtained under the  so-called hypothesis of Minimal Flavour Violation (MFV)~\cite{Chivukula:1987py,Hall:1990ac,D'Ambrosio:2002ex}.
The main idea of MFV is that flavour-violating 
interactions are linked to the
known structure of Yukawa couplings also beyond the SM. 
In a more quantitative way, the MFV construction consists 
in identifying the flavour symmetry and symmetry-breaking structure 
of the SM and enforce it also beyond the SM.

The MFV hypothesis consists of two ingredients: 
(i)~a {\em flavour symmetry} and (ii)~a set of {\em symmetry-breaking 
terms}~\cite{D'Ambrosio:2002ex}.  The symmetry is nothing but the large global 
symmetry ${\mathcal G}_{\rm flavour}$
of the SM Lagrangian in absence of Yukawa couplings
shown in Eq.~(\ref{eq:Gtot}). Since this global symmetry, and particularly 
the ${SU}(3)$ subgroups controlling quark flavour-changing 
transitions, is already broken within the SM, we cannot promote 
it to be an exact symmetry of the NP model. Some breaking would
appear  at the quantum level because of the SM Yukawa interactions.
The most restrictive assumption we can make to {\em protect} in a consistent 
way quark-flavour mixing 
beyond the SM is to assume that $Y_d$ and $Y_u$ are the only 
sources of flavour symmetry  breaking also in the NP model.
If the breaking of the symmetry occurs at very high energy scales, 
at low-energies we would only be sensitive to the background values of 
the $Y$, i.e.~to the ordinary SM Yukawa couplings. 
The role of the Yukawa in breaking the flavour symmetry becomes 
similar to the role of the Higgs in the breaking of the 
gauge symmetry. However, in the case of the Yukawa we don't 
know (and we do not attempt to construct) a dynamical 
model which gives rise to this symmetry breaking.

Within a generic  effective-theory approach to physics beyond the SM, 
we can say that an effective theory satisfies the MFV criterion 
if all higher-dimensional operators,
constructed from SM and $Y_{d,u}$ spurion fields, are invariant under CP and (formally)
under the flavour group ${\mathcal G}_{q}$~\cite{D'Ambrosio:2002ex}.
According to this criterion one should in principle 
consider operators with arbitrary powers of the (dimensionless) 
Yukawa fields. However, a strong simplification arises by the 
observation that all the eigenvalues of the Yukawa matrices 
are small, but for 
the one corresponding to the top quark, and that the off-diagonal 
elements of the CKM matrix are very suppressed. 
Working in the basis in Eq.~(\ref{eq:Ydbasis}) we have 
\be
\left[  Y_u (Y_u)^\dagger \right]^n_{i\not = j} ~\approx~ 
y_t^n V^*_{it} V_{tj}~.
\label{eq:basicspurion}
\ee
As a consequence, in the limit where we neglect light quark masses,
the leading $\Delta F=2$ and $\Delta F=1$ FCNC amplitudes get exactly 
the same CKM suppression as in the SM: 
\begin{eqnarray}
  \mathcal{A}(d^i \to d^j)_{\rm MFV} &=&   (V^*_{ti} V_{tj})^{\phantom{a}} 
 \mathcal{A}^{(\Delta F=1)}_{\rm SM}
\left[ 1 + a_1 \frac{ 16 \pi^2 M^2_W }{ \Lambda^2 } \right]~,
\\
  \mathcal{A}(M_{ij}-\bar M_{ij})_{\rm MFV}  &=&  (V^*_{ti} V_{tj})^2  
 \mathcal{A}^{(\Delta F=2)}_{\rm SM}
\left[ 1 + a_2 \frac{ 16 \pi^2 M^2_W }{ \Lambda^2 } \right]~,
\label{eq:FC}
\end{eqnarray}
where the $\mathcal{A}^{(i)}_{\rm SM}$ are the SM loop amplitudes 
and the $a_i$ are $O(1)$ real parameters. The  $a_i$
depend on the specific operator considered but are flavour 
independent. This implies the same relative correction 
in $s\to d$, $b\to d$, and  $b\to s$ transitions 
of the same type: a key prediction which can be tested 
in experiment.

As pointed out in Ref.~\cite{Buras:2000dm}, within the MFV
framework several of the constraints used to determine the CKM matrix
(and in particular the unitarity triangle) are not affected by NP.
In this framework, NP effects are negligible not only in tree-level
processes but also in a few clean observables sensitive to loop
effects, such as the time-dependent CPV asymmetry in $B_d \to \psi
K_{L,S}$. Indeed the structure of the basic flavour-changing coupling
in Eq.~(\ref{eq:FC}) implies that the weak CPV phase of $B_d$--$\bar
B_d$ mixing is arg[$(V_{td}V_{tb}^*)^2$], exactly as in the SM.  
This construction provides a natural (a posteriori) justification 
of why no NP effects have been observed in
the quark sector, if NP is not far from the TeV scale:  most of the clean observables
measured  so far  are insensitive to NP effects in the MFV
framework. 

\begin{figure*}
\centering
\resizebox{0.45\textwidth}{!}{%
  \includegraphics{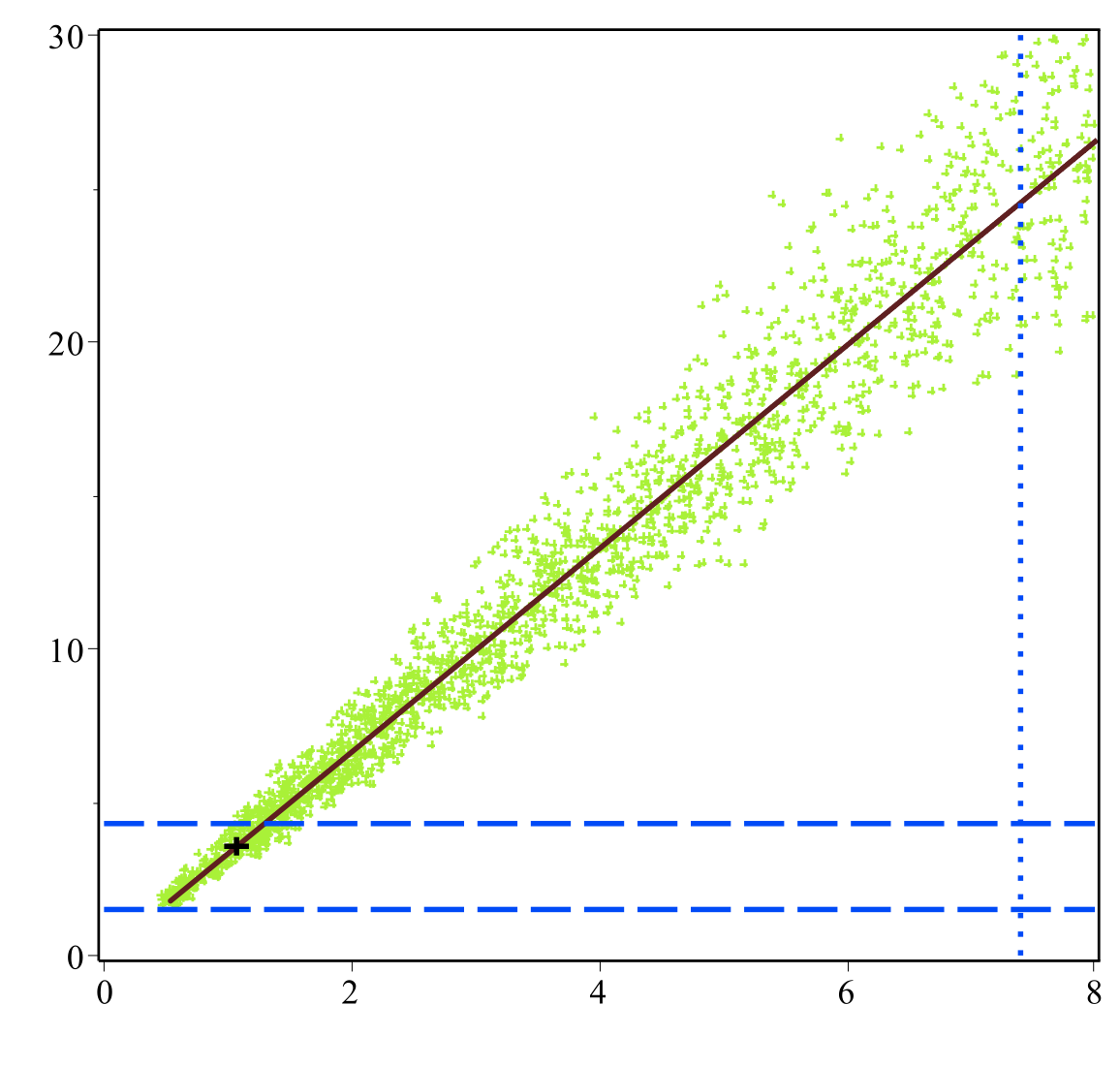}
}
\hskip 0.8 true cm
\resizebox{0.45\textwidth}{!}{%
\includegraphics{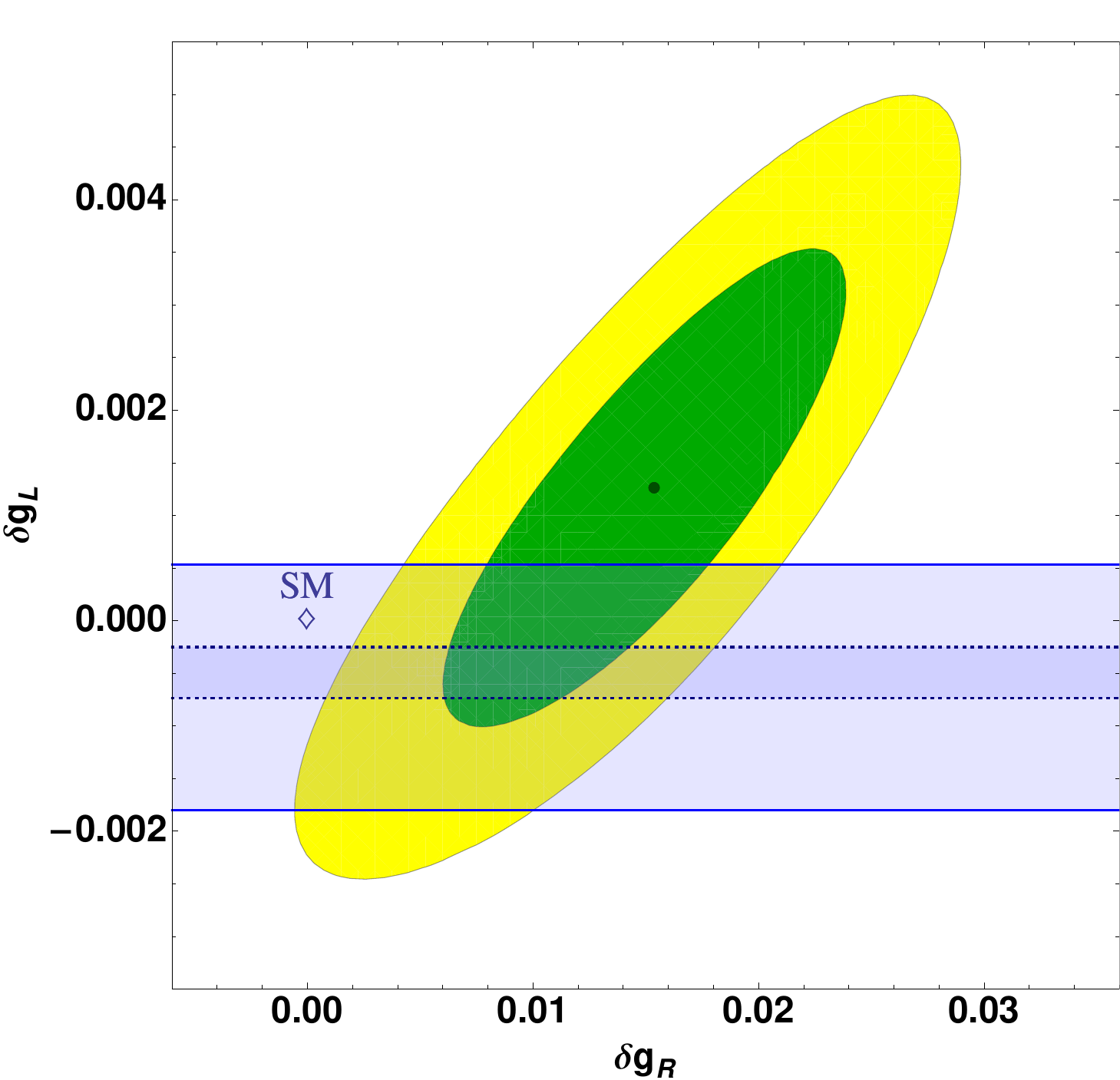}
}
\vskip -7.1 true cm 
\hskip -11  true cm  $10^9 \times \cB(B_{s}\to \mu^+\mu^-)$ \\
\vskip  6.5 true cm
\hskip  -7.5 true cm  $10^{10} \times \cB(B_{d}\to \mu^+\mu^-)$ \\
\vskip  0.2 true cm
\caption{{\em Left:}
Correlation between $\cB(B_s\to\mu^+\mu^-)$ and $\cB(B_d\to\mu^+\mu^-)$
in MFV models. The continuous red line indicates the central value of the correlation, while the 
green points take into account the uncertainties in $|V_{ts}|$ and $|V_{td}|$. 
The black cross denotes the SM prediction. 
The horizontal dashed lines denotes the $95\%$~C.L.~range for $\cB(B_s\to\mu^+\mu^-)$ 
from LHCb and CMS. The vertical dotted line indicates the $95\%$~C.L.~limit on $\cB(B_d\to\mu^+\mu^-)$ 
from LHCb. {\em Right:}  Bounds on possible modified $Z$-boson couplings to down-type quarks from 
flavour-conserving electroweak observables and $\cB(B_s\to\mu^+\mu^-)$,  
assuming MFV~\cite{Guadagnoli:2013mru}.
The inner and outer ellipses denote respectively the 68\% and 95\% 
C.L.~regions as obtained from $Z\to \bar b b$ observables. The horizontal band between full lines 
denote the present 95\% C.L.~constraint from $\cB(B_s\to\mu^+\mu^-)$, while the one
comprised between dotted lines is obtained assuming a future error on 
$\cB(B_s\to\mu^+\mu^-)$ of  about $5\%$.
\label{fig:BsmmNP}}
\end{figure*}

Given the built-in CKM suppression, the bounds on 
higher-dimen\-sio\-nal operators in the MFV framework
turn out to be in the TeV range. 
These bounds are very similar to the 
bounds on flavour-conserving operators derived by precision electroweak tests. 
An illustration of this fact is provided by Fig.~\ref{fig:BsmmNP}~(right), were we 
compare bounds on possible modified $Z$-boson couplings to down-type quarks, under the 
hypothesis of MFV, from 
flavour-conserving electroweak observables and the recent experimental results on $\cB(B_s\to\mu^+\mu^-)$.
This observation reinforces the conclusion that a deeper study of 
rare decays is definitely needed in order to clarify 
the flavour problem: the experimental precision on the clean 
FCNC observables required to obtain bounds more stringent 
than those derived from precision electroweak tests
(and possibly discover new physics) is typically
in the $(1-10)\%$ range.

A few comments are in order: 
\begin{itemize}
\item{}
The MFV ansatz is quite successful on the phenomenological side; however, 
it is unlikely to be an exact property of the model valid to all energy scales. 
Despite some recent attempts to provide a dynamical justification 
of this symmetry-breaking ansatz,
the most natural possibility 
is that MFV is only an accidental low-energy  property of the theory.
Or it could  well  be that a less minimal connection between NP flavour-violating 
couplings and SM Yukawa couplings is at work, as it happens in models of partial 
compositeness (see Sect.~\ref{sect:partial}).
It is then very important to search for possible deviations (even if tiny) 
from the MFV predictions.
\item{}
Even if the MFV ansatz holds, it does not necessarily 
imply small deviations from the SM predictions in all flavour-changing phenomena. 
The MFV ansatz can be implemented in different ways. For instance, in models with 
two Higgs doublets we can change the relative normalization of the two Yukawa 
couplings. It is also possible to decouple the breaking of CP invariance 
from the breaking of the $SU(3)_{Q_L}\times SU(3)_{D_R} \times SU(3)_{U_R}$
quark-flavour group~\cite{Kagan:2009bn}, leaving more room for NP in CP-violating 
observables. All these variations lead to different and well defined patterns 
of possible deviations from the SM that we have only started to investigate.
\item{}
Although MFV seems to be a natural solution to the flavour problem, it should be stressed that  we are still far from having proved the validity of this hypothesis from data.  A proof of the MFV hypothesis can be achieved only with a positive evidence of physics beyond the SM exhibiting the flavour-universality pattern (same relative correction in $s\to d$, $b\to d$, and $b\to s$ transitions of the same type) predicted by the MFV assumption.  While this goal is quite difficult to be achieved, the MFV framework is quite predictive and could easily be falsified. For instance, the rare  modes
 $B_s\to\mu^+\mu^-$ and $B_d\to\mu^+\mu^-$ are strongly correlated in MFV models, as illustrated in Fig.~\ref{fig:BsmmNP}~(left).
 Given no large enhancement over the SM has been observed in $\cB(B_s\to\mu^+\mu^-)$, a possible evidence of $\cB(B_d\to\mu^+\mu^-)$
 well above its SM prediction (perfectly allowed by present data) would  rule out the MFV hypothesis. 
\item{}
The usefulness of the MFV ansatz is closely linked to the theoretical expectation of NP in the TeV range.
This expectation follows from a  {\em natural} stabilization of the Higgs sector, but is disfavored by 
the lack of any direct signal of NP at the LHC. The more the scale of NP is pushed up, the more it is possible to 
allow sizable deviations from the MFV ansatz.
\end{itemize}

\subsection{Explicit examples: I. Supersymmetry}

The Minimal Supersymmetric extension of the SM (MSSM) 
is one of the most studied extensions of the SM at the TeV scale.
Still, despite being "minimal" from the particle point of view,
this model contains a large number of free parameters (especially in the flavour sector)
and we cannot discuss its implications in flavour physics in generality
(namely without specifying in more detail the flavour structure of the model).  
Here we limit ourself to briefly analyze four well-motivated cases:
i) the so-called Constrained MSSM (CMSSM), where the complete model 
is specified in terms of only four free parameters (in addition to the 
SM couplings) defined at some high (grand-unification) scale;
ii)  the NUHM1 scenario: a minimal variation of the CMSSM 
with one extra free parameter allowing non-universal soft masses for the Higgs 
fields, compared to squarks and leptons, at the high scale;
iii) a generic MSSM-type model with heavy first two generations of squarks 
based on the $U(2)^3$ flavour symmetry;
iv) a generic MSSM-type model with all squarks above $\sim 1.5$~TeV but 
for a single light stop.

\begin{figure*}
\centering
\hskip -0.7 true cm
\resizebox{0.55\textwidth}{!}{%
  \includegraphics{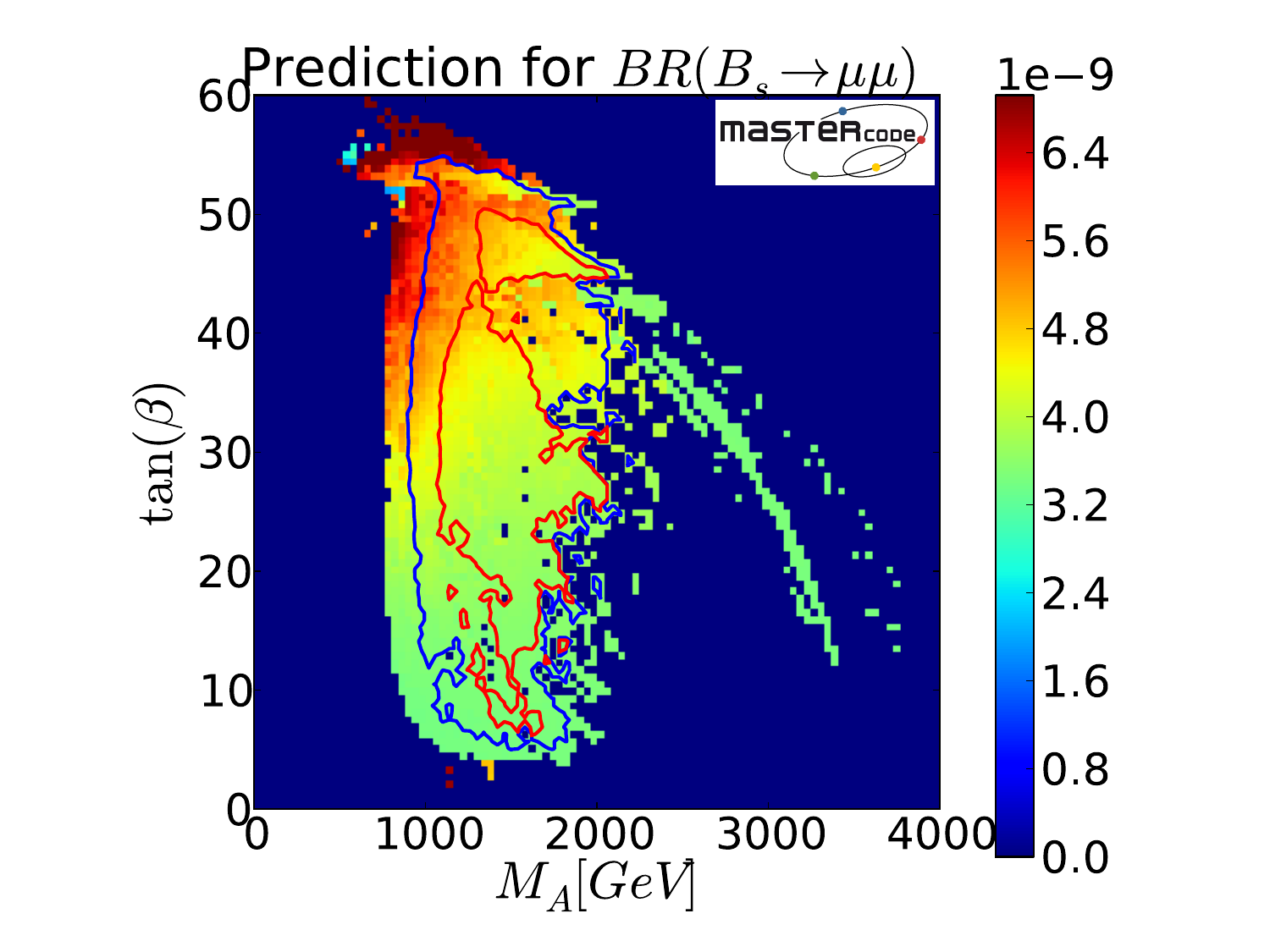}
}
\hskip -1.2 true cm
\resizebox{0.55\textwidth}{!}{%
\includegraphics{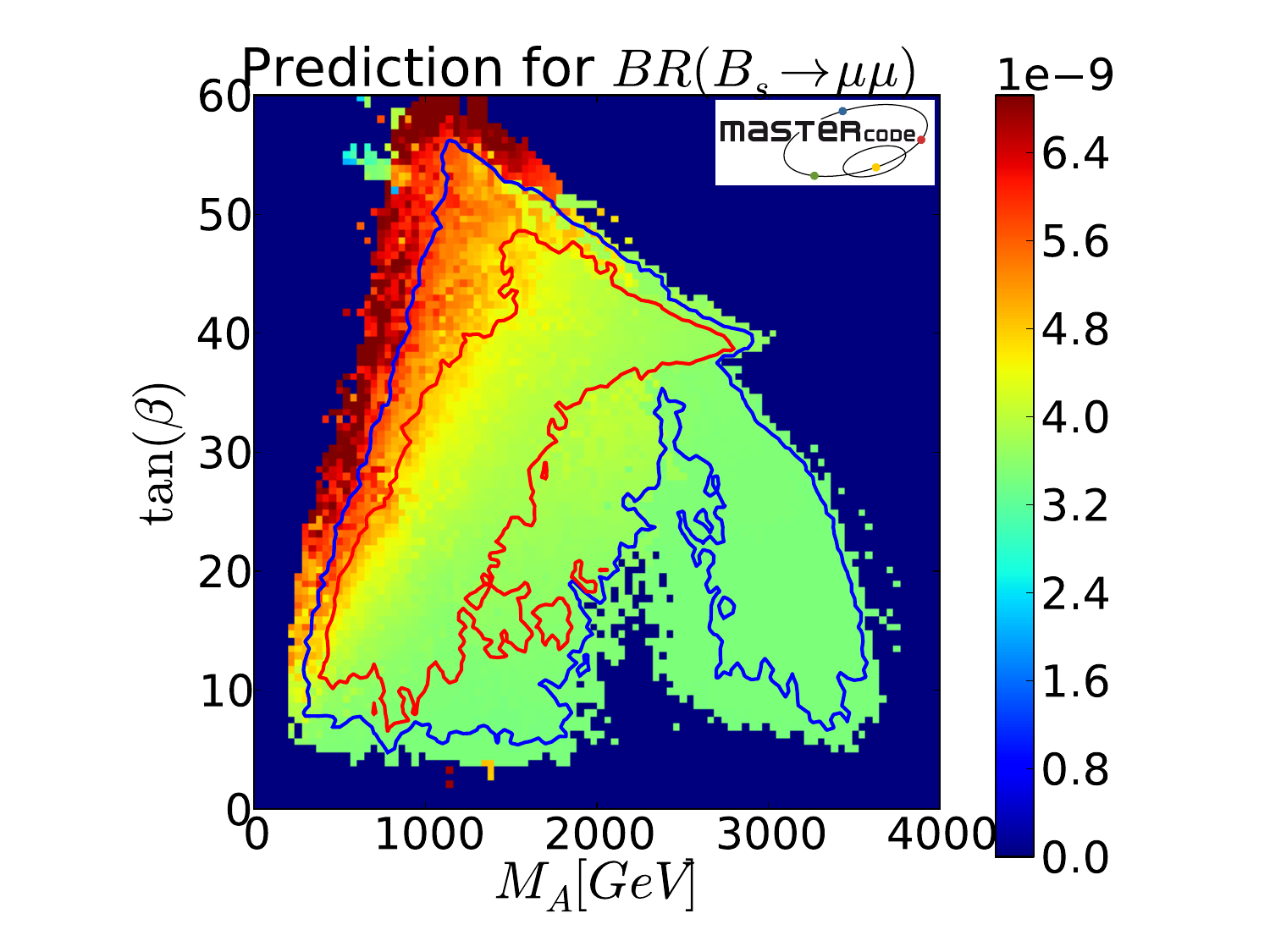}
}
\caption{Predictions for $\cB(B_{s}\to \mu\mu)$ in the 
$M_A$--$\tan \beta$ plane in the CMSSM (left panel) and in the CMSSM
with non-universal Higgs masses (right panel)~\cite{Buchmueller:2012hv}. 
The red and blue contours denote the allowed region of parameter space at 68\% and 98\% C.L.~taking 
into account all LHC data available in autumn 2012 [i.e.~before the experimental evidence of $\cB(B_{s}\to \mu\mu)$]. 
\label{fig:cmssm}}
\end{figure*}

Within the CMSSM and the NUHM1 frameworks the recent experimental data on  $B_s \to \mu^+\mu^-$ do provide 
a very significant constraint on the allowed parameter space (see e.g. Ref.~\cite{Buchmueller:2012hv,Buchmueller:2011sw,Mahmoudi:2012uk}). An illustration of the constraining power of this observable is shown in Fig.~\ref{fig:cmssm}, where we show the prediction of 
 $\BRBsmumu$ in the $M_A$--$\tan\beta$ plane of both models.
As can be seen, the present measurement of $\BRBsmumu$ 
strongly disfavours the region of parameter space with large $\tan\beta$ 
and low $M_A$ values. This constraint is  fully complementary to  
the strong limits already sets on the parameter space of such class of models by the direct searches at ATLAS and CMS. 
It is fair to say that a large fraction of the parameter space of these models 
can be explored only with higher experimental precision on $\BRBsmumu$.
In a long-term perspective, the discovery and the precise measurement of all the 
accessible $B \to \ell^+\ell^-$ channels is one of 
the most interesting items of the $B$-physics program 
at hadron colliders.

The third class of MSSM frameworks we consider is the scenario with heavy first two generations of squarks and with a minimally broken $U(2)^3$ flavour symmetry. As discussed in Ref.~\cite{Barbieri:2011ci,Barbieri:2012uh}, this set up is particularly welcome both to explain why supersymmetry has not been observed yet at the LHC, and also to provide  a natural description of the success of  the CKM picture of flavour mixing and CP violation (beyond MFV). Key low-energy observables in this framework are the  $\Delta F=2$ mixing phases, especially in the $B_{s,d}$ systems. 
A clean prediction is a correlated deviation from the SM in  $B_d$ and $B_s$ mixing, leading to $\Delta_s=\Delta_d =O(10\%) \not=0$.
As shown in Fig.~\ref{fig:bcpv:newphysics}, this possibility is not ruled out yet by present data.  
While it will be very difficult to test this hypothesis in the $B_d$ sector, due to large irreducible 
theoretical uncertainties, there is still significant room for improvements in the $B_s$ system, where the experimental error is still about one order of magnitude larger than the theoretical one. 

The last supersymmetric framework we consider is a MSSM with all squarks above $\sim 1.5$~TeV but 
for a single light stop. This configuration is well consistent with all present collider and flavour data, minimizes the fine-tuning 
problem in the Higgs sector, naturally emerges from the renormalization-group 
evolution of simple UV completions,  and predicts the correct thermal abundance for dark matter~(see e.g.~\cite{Papucci:2011wy,Espinosa:2012in,Delgado:2012eu} and references therein).
Also in this case future improvements in low-energy flavour-physics observables could provide a key information
to test the model. Here the key low-energy flavour observables are $\epsilon_K$ and $B\to X_s \gamma$, 
that are expected to be modified over their corresponding SM
predictions (in a correlated manner) by $(5-10)\%$~\cite{Delgado:2012eu}. 
Observing such a small deviation in the case of $\epsilon_K$ is difficult but not impossible. This would require in particular a
reduction to the few percent level of the error on 
$\rhob$ and $\etab$ via $|V_{ub}|$ and $\gamma$ (see Fig.~\ref{fig:4.1}), that 
represent the dominant (parametric) error in the SM prediction of $\epsilon_K$.

\subsection{Explicit examples: II. Partial compositeness}
\label{sect:partial}

On general grounds, the $\Delta F=1$ and  $\Delta F=2$
flavour-changing operators generated in models with partial compositeness~\cite{Kaplan:1991dc,Agashe:2004cp}  
can be described by the following effective Lagrangians: 
\begin{eqnarray}
\mathcal{L}_{\Delta F=1} & \sim & a^{ab}_{ij} \epsilon_i^{a} \epsilon_j^{b} g_{\rho}
\, \, \frac{v}{m_{\rho}^2} \, \frac{g_{\rho}^2}{(4\pi)^2}
\, \, \overline{\psi}^{a}_i \sigma_{\mu\nu} g_{\rm SM} F_{\rm SM}^{\mu\nu} \psi^{b}_j 
  + 
 b^{ab}_{ij}  \epsilon_i^{a} \epsilon_j^{b} 
\, \, \frac{g_{\rho}^2}{m_{\rho}^2} 
\, \, \overline{\psi}^{a}_i \gamma^{\mu} \psi^{b}_j i H^{\dagger} \overleftrightarrow{D}_{\mu} H  \nonumber \\
\mathcal{L}_{\Delta F=2} & \sim &
c^{abcd}_{ijkl} \epsilon_i^{a} \epsilon_j^{b}  \epsilon_k^{c} \epsilon_l^{d} 
\, \,  \frac{g_{\rho}^2}{m_{\rho}^2}
\, \, \overline{\psi}^{a}_i \gamma^{\mu} \psi^{b}_j \, \overline{\psi}^{c}_k \gamma_{\mu} \psi^{d}_l  \label{eq:LagrangianPC}~.
\end{eqnarray}
Here $g_\rho \lesssim 4\pi$ and $m_\rho$ denote the 
coupling constant and the mass scale of the resonances in the composite sector, while
$g_{\rm SM}$ and $F_{\rm SM}^{\mu\nu}$
denote generically couplings and field strengths of the SM gauge fields. The 
$a^{ab}_{ij}$,  $b^{ab}_{ij}$,  and $c^{abcd}_{ijkl}$ are numerical coefficients, 
depending on the details of the strong dynamics.
Finally, the $\epsilon_i^{a}$ are the (flavour-dependent)
parameters controlling the mixing of the elementary fermions with heavy fermonic resonances having the same electroweak quantum 
numbers (see \cite{Davidson:2007si,KerenZur:2012fr} fore more details).

In first approximation,  the elementary fermions can be identified with the  SM fermions, that we label as 
$\psi^{a}_i$  (where $a=Q,U,D,L,E$ denotes the fermion species and $i=1\ldots3$ the flavour index, see sect.~\ref{sec:2}). 
In this framework, the Yukawa couplings originate from the 
mixing between elementary and composite fermions, and can be decomposed as follows
\begin{equation}  \label{eq:yukawas}
(Y_u)_{ij} \sim g_{\rho} \epsilon^q _i \epsilon^u_j
\quad , \quad
(Y_d)_{ij} \sim g_{\rho} \epsilon^q _i \epsilon^d_j 
\quad , \quad
(Y_e)_{ij} \sim g_{\rho} \epsilon^\ell _i \epsilon^e_j \, ,
\end{equation}
where an $O(1)$ coefficient is understood for each entry of the $Y$'s. As can be seen, in this class of models the fundamental parameters
controlling flavour-breaking effects are the $\epsilon^e_j $ and not the Yukawa couplings (as in MFV). However, $\epsilon^e_j $ and Yukawa couplings
are connected ensuring a sufficient protection of flavour-mixing effects involving light generations.

In the quark sector, going to the CKM basis and requiring the quark masses and the mixing angles to be naturally reproduced, one is left with 2 free parameters (e.g. $g_\rho$ and $\epsilon^u_3$). The parameters $\epsilon_i^a$ represent the ``degree of partial compositeness'' of the various fields. Generally speaking the electroweak precision tests 
(EWPT) are more easily satisfied if the SM fields are mostly elementary, that is if the  $\epsilon_i^a$ are small and $g_\rho \sim 4\pi$.

In the case of composite Higgs models, one expects $a^{ab}_{ij}$,~$b^{ab}_{ij}$,~$c^{abcd}_{ijkl}=O(1)$, and this framework is compatible with the strongest flavour bounds in kaon physics provided that $m_\rho \gtrsim 10$ TeV. As shown in the recent analysis of Ref.~\cite{KerenZur:2012fr},
in this case  one can expect deviations from the SM at the present level of experimental sensitivity 
 in the  electric dipole moment (EDM) of the neutron (where there is actually a significant tension with the present bound),  
 CP-violating observables in the kaon system ($\epsilon'/\epsilon$ and $\epsilon_K$),
 and $b\to s$ FCNC transitions. However, in the lepton sector
 the minimal framework is not satisfactory (a severe fine-tuning is needed to satisfy current bounds on lepton-flavour violating processes).
This general construction provides an effective description of a wide class of partially-composite 
models. However, it should be stressed that also in  partial-composite models it is possible to postulate the existence of additional 
protective flavour symmetries (as discussed for instance in Ref.~\cite{Redi:2011zi,Barbieri:2012bh,Barbieri:2012tu}) and, for instance, recover a MFV structure. In this case the bounds on $m_\rho$ from flavour constraints are well below $10$ TeV.

The general flavour structure of partial compositeness can also be realized in the context of Supersymmetry~\cite{Nomura:2007ap}. In this case the various coefficients are computable in terms of the supersymmetry breaking parameters and are loop suppressed. 
Interestingly, in this case quarks masses $\tilde{m}=O(\mbox{1 TeV})$ are consistent with flavour bounds both in the quark and in the lepton sector. The most promising observables in which sizable deviations from the SM can appear are again the neutron EDM (with reduced tension) and $\epsilon'/\epsilon$, together with the electron EDM and the $\mu \rightarrow e\gamma$ transition~\cite{KerenZur:2012fr}.

It is worth to stress that in both frameworks (with or without supersymmetry, in absence of additional flavour symmetries), with $m_\rho \sim 10$ TeV or $\tilde{m}\sim 1$ TeV, 
it is easy to generate direct CP violating asymmetries in D-meson decays at the ${\rm few}\times0.1\%$ level. Moreover, 
deviations from the SM at the $(10-20)\%$ level can show up in the process $K^+\to\pi^+\bar\nu\nu$,  possibly within the reach of the planned sensitivity of the NA62 experiment~\cite{NA62_future}.

\section{Conclusions}

As we have discussed in general terms and with a few explicit examples, flavour physics is a key tool to investigate the nature of physics beyond the SM. The recent discovery of a new state with mass around 125 GeV, compatible with properties of the SM Higgs boson
(and pointing toward the existence of a fundamental Higgs field),
makes the case of future high-precision studies in flavour physics even more motivated: all the key properties of low-energy 
flavour physics are determined by the Yukawa couplings, or by the couplings of the Higgs field to the fermions. 
As we have seen, in several cases our knowledge of the Yukawa sector is still quite limited (often not exceeding the $20\%$ relative 
accuracy for amplitudes forbidden at the tree level).  A deeper investigation of flavour physics is therefore a necessary element for a deeper understanding of the properties of the Higgs field.

More generally, flavour physics has a twofold role in investigating the nature of physics beyond the SM. 
On the one hand, for NP modes with new particles close to the TeV scale, existing low-energy 
flavour-physics bounds put very stringent limits on the flavour structure of the model.
Present data already tell us that the new degrees of freedom must have a highly non-trivial 
flavour structure (MFV-like) in order to be consistent with observations.
In this perspective, if direct signals of NP will appear during the next LHC run, 
future progress in flavour physics will be an essential  tool to better investigate 
the peculiar flavour structure of the new degrees of freedom.

On the other hand, the paradigm of NP at the TeV scale is seriously challenged by the absence of 
deviations from the SM at the high-energy frontier. In this perspective, flavour physics remains a very powerful
tool to search for physics beyond the SM, being potentially sensitive to NP scales much higher than 
those directly accessible at present and near-future high-energy facilities, for models with a generic flavour 
structure. 

Making progress in this field is mainly a question of precision,
both on the theory and on the experimental side: visible deviations from the SM may be  at the origin  
of some of the existing "tensions" between data and SM predictions, or may simply be around the corner in terms of statistical 
precision. The key point is to identify observables sensitive to short-distance physics whose 
theory uncertainty is sufficiently under control, in order to perform more sensitive tests of the model with the help of 
more accurate experimental data.

%
%
%

%
%

\end{document}